\documentclass[12pt]{article}
\usepackage{amsfonts}
\usepackage{amsthm}
\input{epsf.sty}
\usepackage{latexsym,amsmath,amsfonts,amscd}
\usepackage{epsfig}
\usepackage{changebar}
\usepackage{pstricks}
\usepackage{pst-plot}
\usepackage{multirow}
\usepackage{subfigure}
\usepackage{placeins}
\usepackage{amssymb}
\usepackage{mathrsfs}
\usepackage[title]{appendix}
\usepackage{bm}
\usepackage{hyperref}
\usepackage{stackengine}
\usepackage[affil-it]{authblk} 

\usepackage{multirow,bigstrut}
\usepackage{enumitem}
\usepackage{booktabs}

\numberwithin{equation}{section}

\theoremstyle{plain}
\newtheorem{thm}{Theorem}[section]

\newtheorem{rem}[thm]{Remark}

\theoremstyle{definition}

\newcommand{\brac}[1]{\left(#1\right)}
\newcommand{\abs}[1]{\left\vert#1\right\vert}

\begin{document}

% \font\myfont=cmr12 at 40pt

\baselineskip=1.5pc
% \vspace{.5in}

\begin{center}

{\large\bf Learning Thermodynamically Stable and Galilean Invariant Partial Differential Equations for Non-equilibrium Flows}

% \title{Learning Interpretable and Thermodynamically Stable Partial Differential Equations}

\end{center}

\vspace{.1in}

\centerline{
Juntao Huang \footnote{Department of Mathematics,
Michigan State University, East Lansing, MI 48824, USA.} \quad
%E-mail: huangj75@msu.edu. 
Zhiting Ma \footnote{Department of Mathematical Sciences, 
Tsinghua University, Beijing, China.}\quad
%E-mail: liu@math.wichita.edu.
%Research supported in part by a grant from the Simons Foundation (426993, Yuan Liu). 
%Yizhou Zhou \footnote{Department of Mathematical Sciences, 
%Tsinghua University, Beijing, China.}\quad
Yizhou Zhou \footnotemark[2]\quad
Wen-An Yong \footnotemark[2]\textsuperscript{,} 
\renewcommand{\thefootnote}{\fnsymbol{footnote}}
\footnote{E-mail: wayong@tsinghua.edu.cn. Corresponding author}
%Wen-An Yong \footnote{Department of Mathematical Sciences, 
%Tsinghua University, Beijing, China. E-mail: wayong@tsinghua.edu.cn. Corresponding author}
}

% \author{Juntao Huang\thanks{Department of Mathematics,
% Michigan State University, East Lansing, MI 48824, USA.} \and
% Zhiting Ma \thanks{Department of Mathematical Sciences, 
% Tsinghua University, Beijing, China.} \and
% Yizhou Zhou \footnotemark[2]\and
% Wen-An Yong \thanks{Corresponding author} \footnotemark[2]
% }

% \maketitle

\vspace{.4in}

% \begin{abstract}
\centerline{\bf Abstract}

\vspace{.1in}

In this work, we develop a method for learning interpretable, thermodynamically stable and Galilean invariant partial differential equations (PDEs) based on the Conservation-dissipation Formalism of irreversible thermodynamics. As governing equations for non-equilibrium flows in one dimension, the learned PDEs are parameterized by fully-connected neural networks and satisfy the conservation-dissipation principle automatically. In particular, they are hyperbolic balance laws and Galilean invariant. The training data are generated from a kinetic model with smooth initial data. Numerical results indicate that the learned PDEs can achieve good accuracy in a wide range of Knudsen numbers. Remarkably, the learned dynamics can give satisfactory results with randomly sampled discontinuous initial data and Sod's shock tube problem although it is trained only with smooth initial data.

% \end{abstract}
\vspace{.1in}

% \bigskip

% \bigskip
% \vfill


%{\bf AMS(MOS) subject classification:} 65M99

\pagenumbering{arabic}

%section 1
\section{Introduction}\label{intro}

\setcounter{equation}{0}
\setcounter{figure}{0}
\setcounter{table}{0}

Emergence of machine learning and data science has brought new vitality to the development of 
various scientific disciplines. Mostly, such disciplines are based on certain fundamental principles which can be described with mathematical frameworks. The frameworks usually contain problem-dependent freedoms. Besides the traditional approaches like experiments, machine learning is a promising alternative in fixing the freedoms. In this sense, the booming data-based technique can display its power by a proper combination with reliable scientific theories.

Thermodynamics is a classical theory studying relationships among various apparently unrelated variables or parameters characterizing a thermodynamic system. 
For non-equilibrium systems, it was explained in \cite{Yong2020} that the most natural form of the expected relations are evolution partial differential equations (PDEs).
Thus, the core task of irreversible thermodynamics is choosing suitable variables and determining their evolution equations. 
However, no well-accepted rules have been given so far and consequently there are many different schools,
such as Classical Irreversible Thermodynamics \cite{De2013}, 
Rational Extended Thermodynamics \cite{muller1998rational}, Extended Irreversible Thermodynamics \cite{jou1996extended,lebon2008understanding}, General Equation for Non-equilibrium Reversible–irreversible Coupling \cite{ottinger2005beyond,pavelka2018multiscale},
Conservation-dissipation Formalism (CDF) \cite{Zhu2015}, Energetic Variational Approach \cite{hyon2010energetic}, and so on. 
An exposition of different versions of thermodynamics can be found in \cite{muschik2014contact} and references cited therein.

This project focuses on a combination of machine learning with the CDF theory mentioned above.
CDF is based on the first and second laws of thermodynamics. It assumes that certain conservation laws are known a priori. By postulating the existence of an entropy function and a dissipative matrix, we follow a Conservation-dissipation Principle distilled in \cite{Yong2008} to choose the non-equilibrium variables and derive the corresponding evolution equations. 
The PDEs thus obtained are hyperbolic balance laws (first-order PDEs) which naturally meet four fundamental physical requirements proposed in \cite{Yong2020}: the observability of physical phenomena, time-irreversibility, long-time tendency to equilibrium, and compatibility with existing classical theories. The importance of these requirements were expounded in \cite{Yong2020}. 
In CDF, the entropy function and dissipative matrix are problem-dependent and constitute its freedoms.
These freedoms can be fixed by resorting to machine learning. 
Once the freedoms are determined, we obtain interpretable and thermodynamically stable PDEs governing the thermodynamic system under consideration.

As a start of this project, the present paper is devoted to finding  the governing macroscopic equations for one-dimensional non-equilibrium flows. 
To be specific, we assume that the flow obeys the mesoscopic BGK (Bhatnagar-Gross-Krook) equation in kinetic theory. Thus, the classical conservation laws of mass, momentum and energy are known a priori. By adding one more (non-equilibrium) variable, we obtain a hyperbolic system of first-order PDEs with four unknown variables, which is Galilean invariant. Then we use fully connected neural networks to represent the entropy function and dissipation matrix. 
The training data are generated by numerically solving the BGK model with smooth initial data. The neural network representations for the hidden dynamics are trained by minimizing a loss function based on a discrete version of the CDF-based PDEs acting on the training data. 

Numerical results show that our CDF-based machine learning model indeed satisfies the conservation-dissipation principle. This particularly implies that the learned PDEs are symmetrizable hyperbolic. {Moreover, the model can achieve good accuracy in a wide range of Knudsen numbers for both randomly sampled smooth and discontinuous initial data. In particular, its performance on Sod's shock tube problem \cite{sod1978survey} is better than the Euler equations in both hydrodynamic and kinetic regimes. Remarkably, the learned dynamics can give satisfactory results with discontinuous initial data although it is trained only with smooth initial data.}

The macroscopic governing equations obtained above can also be understood as a moment closure system of the BGK model. However, it is important to point out that the CDF does not necessarily rely on the kinetic theory. {Here we use the kinetic model just because we have no other means to obtain the training data.} We are expecting to have such data from experiments or other means so that the CDF framework can exhibit its potential in a wide range of problems. 
In addition, note that the CDF theory itself allows more non-equilibrium variables. Here we choose only one non-equilibrium variable merely for the sake of simplicity. It is possible to achieve better results by adding more non-equilibrium variables.

As to the moment closure system, we mention the recent work \cite{han2019uniformly}, where the authors introduced a framework to construct interpretable moment closure models for kinetic problems. They firstly learned a set of generalized moments using the auto-encoder to optimally represent the underlying velocity distribution, and then learned the moment closure model for the generalized moments using neural networks with the aim of best capturing the associated dynamics of the kinetic equation. In this way, some novel moment closure systems are obtained and give good prediction results. 
Nevertheless, it is not clear whether the moment closure systems obtained in \cite{han2019uniformly} are hyperbolic or preserve the dissipation property of the kinetic equation characterized by the H-theorem. In contrast, the hyperbolicity and dissipation property are enforced in our CDF-based PDEs automatically. 

Other recent works on learning PDEs have the following.  
In \cite{ling2016reynolds}, the authors designed a neural network with Galilean invariance to discover the Reynolds stress anisotropy tensor in 
turbulence models from high-fidelity simulation data. 
In \cite{lei2020machine}, a neural network which preserves the rotational symmetry was constructed to approximate the constitutive laws in non-Newtonian fluid models. 
In \cite{raissi2018hidden,raissi2019physics}, by resorting to the Gaussian process regression or neural networks named Physics-Informed Neural Networks, some classical PDEs with unknown parameters or functions are explored. This was further applied in many realistic problems \cite{chen2020solving,zhang2020physics,yin2020non}.
In \cite{brunton2016discovering,rudy2019data}, a data-based method called SINDy (Sparse Identification of Nonlinear Dynamics) was developed to discover dynamical systems from the perspective of sparse regression and compressed sensing. 
In \cite{long2018pde1,long2019pde2}, the authors used a symbolic neural network called PDE-Net to uncover the analytic form of PDEs.

There are more works about learning ordinary differential equations (ODEs). For Hamiltonian systems in classical mechanics, different neural networks such as SympNets and H\'enonNet have been designed in \cite{jin2020symplectic,burby2020fast}. More works on learning Hamiltonian systems can be found in \cite{greydanus2019hamiltonian,toth2019hamiltonian,zhong2019symplectic} and references cited therein.
In \cite{kolter2019learning}, ODEs with dissipation structure are considered and the dissipation structure 
was built into the neural network by jointly learning the dynamic model and a Lyapunov function.
In a very recent work \cite{yu2020onsagernet}, a generalized Onsager principle was proposed  
and a so-called OnsagerNet was designed.
In a series of works \cite{han2017deep,zhang2018deep,wang2018deepmd,zhang2018end}, deep neural networks with symmetry-preserving property are explored to present the potential energy surface (PES) that describes the interaction between all the nuclei in the system in the molecular dynamics. 
For more details of the development of the physics-based modeling with machine learning, we refer readers to the review paper \cite{han2020integrating}.

The paper is organized as follows. The general CDF theory is briefly reviewed in Section \ref{sec:preliminary}. In Section \ref{section3}, we construct a simple CDF-based model for non-equilibrium flows. Section \ref{section4} is devoted to a detailed procedure to learn the freedoms in the model.
The effectiveness of the procedure is demonstrated through numerical results in Section \ref{sec:numerical}. Some concluding remarks are given in Section \ref{sec:conclusion}.

\section{Preliminaries}\label{sec:preliminary}
In this section, we briefly review the CDF theory \cite{Zhu2015}.
It is concerned with an irreversible process which obeys some conservation laws 
of the form
\begin{equation}\label{conservation-laws}
	\partial_t u + \sum_{j=1}^d \partial_{x_j}f_j = 0.
\end{equation}	
Here $u = u(t, x) \in \mathbb{R}^n$ represents conserved variables depending on the time and spatial coordinates $(t, x)$, $x \in \mathbb{R}^d$ with $d$ the dimension of space, and $f_j$ is the corresponding flux along the $x_j$-direction. 
If each $f_j$ is given in terms of the conserved variables, the system \eqref{conservation-laws} becomes closed. In this case, the system is considered to be in local equilibrium and $u$ is also referred to as equilibrium variables. 
However, very often $f_j$ depends on some extra variables in addition to the conserved ones.
The extra variables characterize non-equilibrium features of the system under consideration, called non-equilibrium or dissipative variables, and their choice is not unique.

Motivated by the Extended Thermodynamics \cite{muller1998rational,jou1996extended}, 
in CDF we choose a set of dissipative variables $v\in \mathbb{R}^m$ so that the flux $f_j$ in \eqref{conservation-laws} can be expressed as $f_j = f_j(u, v)$ and seek evolution equations of $v$ in the form
\begin{equation}\label{2.2}
	\partial_t v + \sum_{j=1}^d \partial_{x_j} g_j(u, v) = q(u, v).
\end{equation}
This is our constitutive equation. 
Here $g_j(u, v)$ is the corresponding flux and $q = q(u, v)$ is the source vanishing at equilibrium.
Together with the conservation laws \eqref{conservation-laws}, the evolution of the non-equilibrium system is governed by first-order PDEs 
\begin{equation}\label{CDF-PDE}
	\partial_t U + \sum_{j=1}^d \partial_{x_j} F_{j}(U) = Q(U),
\end{equation}
where
\begin{equation*}
	\begin{aligned}
		U = \begin{pmatrix}
			u \\[2mm] 
			v
		\end{pmatrix}, \qquad
		F_j(U) = \begin{pmatrix}
			f_j(U) \\[2mm] 
			g_j(U)
		\end{pmatrix}, \qquad
		Q(U) = \begin{pmatrix}
			0 \\[2mm] 
			q(U)
		\end{pmatrix}.
	\end{aligned}
\end{equation*}

It was observed in \cite{Yong2008} that many classical PDEs of the form \eqref{CDF-PDE} respect the following so-called Conservation-dissipation Principle:
\begin{enumerate}
	\item[(i)] There is a strictly concave smooth function $\eta = \eta(U)$, called entropy (density), such that the matrix product $\eta_{UU}F_{jU}$ is symmetric for each $j$ and for all $U$ under consideration. \label{condition1}
	\item[(ii)] There is a positive definite matrix $M = M(U)$, called dissipation matrix, such that the non-zero source can be written as $q(U) = M(U)\eta_v(U)$.\label{condition2}
\end{enumerate}
Here the subscript stands for the corresponding partial derivative, for instance $\eta_v = \partial \eta/\partial v$ and $\eta_{UU} = \partial^2 \eta/\partial U^2$, and $\eta_v(U)$ should be understood as a column vector. Note that the dissipation matrix is not assumed to be symmetric, and its positive definiteness means that of the symmetric part $(M + M^T)/2$. 

Inspired by the above observation, we try to determine the constitutive relation \eqref{2.2} so that the resultant balance laws \eqref{CDF-PDE} also respect the conservation-dissipation principle (see an example in the next section). This is our Conservation-dissipation Formalism (CDF) proposed in \cite{Zhu2015}.

This formalism has two freedoms: the entropy function $\eta =\eta(U)$ and the dissipation matrix
$M=M(U)$. They are both functions of the state variable $U$. The former is strictly concave and the latter is positive definite. Except these, no further restriction is imposed on the freedoms. Specific expressions of $\eta(U)$ and $M(U)$ should be problem-dependent. 
With these freedoms, CDF provides a tailored platform for machine learning to display its power.

On the conservation-dissipation principle, we make the following simple comments. Condition (i) is the well-known entropy condition for hyperbolic conservation laws \cite{Friedrichs1971,Godunov1961}.
It ensures that the first-order system \eqref{CDF-PDE} is globally symmetrizable hyperbolic and thereby well-posed \cite{Dafermos2005}. 
Condition (ii) is a nonlinearization of the celebrated Onsager reciprocal relation for scalar irreversible processes \cite{De2013}. Together with the first condition, it can be regarded as a stability criterion for non-equilibrium thermodynamics. 
As shown in \cite{Yong2020}, this criterion guarantees that the CDF-based PDEs reflect the observability of physical phenomena, time-irreversibility, long-time tendency to equilibrium, and compatibility with existing classical theories.
Further comments can be founded in \cite{Yong2020}.

\section{A CDF-based model}\label{section3}

Consider an electrically neutral fluid without external forces. Its motion obeys the conservation laws of mass, momentum and energy \cite{jou1996extended}:
\begin{equation}\label{eq:conservation-laws}
	\begin{aligned}
		\partial_t \rho + \nabla \cdot (\rho v) &= 0,\\
		\partial_t (\rho v) + \nabla \cdot(\rho v \otimes v + {\bf P}) &= 0,\\
		\partial_t E+ \nabla \cdot(E v + q + {\bf P} v) &= 0.\\
	\end{aligned}
\end{equation}
Here $\rho$ is the fluid density, $v\in \mathbb{R}^d$ is the velocity, $E =\rho e + \frac{1}{2}\rho |v|^2$ is the total energy with $e$ the specific internal energy, ${\bf P}$ is the pressure tensor, and $q$ represents the heat flux. This system gives the evolution of variables $(\rho, \rho v, E)$. 
It is closed only if ${\bf P}$ and $q$ are specified. In \cite{Zhu2015}, the CDF was used to obtain a set of evolution equations for these variables with the two freedoms--- the corresponding entropy function and the dissipation matrix. 

The goal of this paper is to fix the freedoms with machine learning. For simplicity, we only consider the one-dimensional problem. Moreover, in order to have reasonable training data, we turn to the one-dimensional BGK model  
\begin{equation}\label{eq:kinetic-bgk}
	\frac{\partial f}{\partial t} + \xi \frac{\partial f}{\partial x} = \frac{1}{\varepsilon}(f_M - f),
\end{equation}
meaning that the fluid flow is governed by this kinetic model.
Here $f=f(x,t,\xi)$ is a distribution function with $\xi\in \mathbb{R}$ the particle velocity, $\varepsilon$ is the dimensionless Knudsen number, and $f_M$ is the Maxwellian
\begin{equation*}
	f_M=f_M(\xi; \rho, v, T) = \frac{\rho}{(2\pi T)^{1/2}}\exp\brac{-\frac{(\xi-v)^2}{2T}}
\end{equation*}
with  
\begin{equation}\label{rhovT}
	\begin{aligned}
	& \rho = \int_{\mathbb{R}} f d\xi, \quad \rho v = \int_{\mathbb{R}} \xi f d\xi,  
	& \rho T = \int_{\mathbb{R}} (\xi-v)^2 f  d\xi.
	\end{aligned}
\end{equation}

It is well-known that the one-dimensional BGK model implies the conservation laws \eqref{eq:conservation-laws} with the pressure tensor ${\bf P}$ becoming a scalar $p = \rho T$, $e = T/2$, 
\begin{equation}\label{E and q}
	\begin{aligned}
	E = \frac{1}{2} \int_{\mathbb{R}} \xi^2 f  d\xi = \rho e + \frac{1}{2}\rho v^2, \qquad
	q = \frac{1}{2}\int_{\mathbb{R}}  (\xi-v)^2(\xi-v) f d\xi. 
	\end{aligned}
\end{equation}
In fact, multiplying \eqref{eq:kinetic-bgk} with $1, \xi, \frac{1}{2}\xi^2$ and integrating the resultant equations over $\xi\in \mathbb{R}$ will immediately lead to the conservation laws \eqref{eq:conservation-laws} (see e.g. \cite{struchtrup2005macroscopic}). Notice that for the one-dimensional kinetic model, we have the following equation of state $p=2\rho e$.
Thus, only the heat flux $q$ in \eqref{eq:conservation-laws} is an extra variable.

To close \eqref{eq:conservation-laws} with CDF, we firstly recall that the system in equilibrium has a specific entropy \cite{struchtrup2005macroscopic}
\begin{equation*} 
	s^{\textrm{(eq)}} = s^{\textrm{(eq)}}(\nu, e)=- k_b\nu \int_{\mathbb{R}} f_M \ln f_M d \xi = k_b\left(\frac{1}{2} \ln e + \ln \nu\right) + C,
\end{equation*}
where $k_b$ is the Boltzmann constant, $\nu = 1/\rho$ is the specific volume, and $C$ is a constant. Clearly, this entropy function is concave with respect to $(e,\nu)$. In what follows, we take $k_b=1$ for simplicity.

Furthermore, we introduce a new dissipative variable $w$ and postulate that a specific entropy of the form 
\begin{equation*} 
s = s(\nu, e, w; \varepsilon) = s^{\textrm{(eq)}}(\nu, e) + s^{\textrm{(neq)}}(w ;\varepsilon)
\end{equation*}
exists for the non-equilibrium system under consideration.  
Here for simplicity we have assumed that there is no cross-terms between $w$ and $(\nu, e)$. 
According to the conservation-dissipation principle, the undetermined and non-equilibrium part $s^{\textrm{(neq)}}(w;\varepsilon)$ should be a concave function of $w$ with the Knudsen number $\varepsilon$ as a parameter.  
For such a concave function $s = s(\nu, e, w; \varepsilon)$, it is well-known that the function $\eta$ defined as 
\begin{equation*}
	\eta = \eta(\rho, \rho v, E, \rho w;\varepsilon) = \rho s(\nu, e, w;\varepsilon)
\end{equation*}
can act as the entropy in the conservation-dissipation principle. Namely, it is strictly concave with respect to $(\rho, \rho v, E, \rho w)$.  

Next we follow \cite{Zhu2015} and introduce the differential operator $D$ acting on a function $h=h(x, t)$ as $Dh := \partial_t (\rho h) + \partial_x (\rho v h)=\rho (\partial_t h + v \partial_x h)$. Using the conservation laws \eqref{eq:conservation-laws}, we calculate 
\begin{equation*} 
	\begin{aligned}
		\eta_t+\partial_x(v \eta)\equiv Ds&=s_\nu D\nu + s_e D e + s_w D w \\
		& = \rho \partial_x v + \theta^{-1}(-\partial_x q- \rho\theta \partial_x v) + s_w D w  \\
		&=-\theta^{-1}\partial_x q +s_w Dw\\
		&= -\partial_x (\theta^{-1}q) + s_w Dw + q \partial_x \theta^{-1} 
	\end{aligned}
\end{equation*}
with $\theta^{-1}:= s_e$. 
Here we have used $s_{\nu}=\rho$ and $s_e = 1/(2e)$. 
This calculation suggests that $\theta^{-1}q$ is the entropy flux and $s_w Dw + q \partial_x \theta^{-1}$ is the entropy production. 

According to the calculation above, we choose the heat flux 
$$
q = s_w = s^{\textrm{(neq)}}_w(w;\varepsilon)
$$ 
and the evolution equation for $w$ as 
\begin{equation*}
	\partial_t(\rho w) + \partial_x(\rho v w) + \partial_x \theta^{-1} = M q 
\end{equation*}
with {$M=M(\rho, e, w; \varepsilon)$} a positive function. 
{Here $M$ is independent of the velocity $v$ such that the model is Galilean invariant.}
By such a choice, we have the entropy production 
$$
s_w Dw + q \partial_x \theta^{-1}=q Mq\geq 0. 
$$
Moreover, we notice that the non-equilibrium variable $w$ can be globally expressed in terms of $q$ and $\varepsilon$ due to the strict concavity of $s^{\textrm{(neq)}}(w ;\varepsilon)$ with respect to $w$. 

Consequently, we arrive at the following balance laws
\begin{equation}\label{eq:CDF-w}
	\begin{aligned}
		\partial_t \rho + \partial_x (\rho v) &= 0,\\
		\partial_t (\rho v) + \partial_x(\rho v^2 + \rho \theta ) &= 0,\\
		\partial_t E + \partial_x(E v + s^{\textrm{(neq)}}_w + \rho \theta v) &= 0,\\
		\partial_t(\rho w) + \partial_x(\rho v w) +\partial_x \theta^{-1} &= M s^{\textrm{(neq)}}_w
	\end{aligned}
\end{equation}
with the freedoms $M=M(\rho, e, w;\varepsilon)$ and $s^{\textrm{(neq)}}=s^{\textrm{(neq)}}(w;\varepsilon)$. Here we notice that the temperature $T=\theta$ for the one-dimensional kinetic model. 
{This system satisfies the conservation-dissipation principle and thereby is globally symmetrizable hyperbolic. Moreover, it is Galilean invariant.}

\section{Learning the freedoms}\label{section4}

In this section, we present a procedure to learn the two undetermined functions $s^{\textrm{(neq)}}=s^{\textrm{(neq)}}(w;\varepsilon)$ and $M=M(\rho, e, w;\varepsilon)$ in \eqref{eq:CDF-w}.  
Note that an unlimited amount of training data $(\rho, \rho v, E, q)$ can be obtained by numerically solving the BGK model and then by computing the moments in \eqref{rhovT}-\eqref{E and q}. The details are given in Section \ref{sec:numerical}.

Although the training data $(\rho, \rho v, E, q)$ are available, the values of $w$ cannot be obtained directly since $s^{\textrm{(neq)}}=s^{\textrm{(neq)}}(w;\varepsilon)$ has not been fixed yet.  
Therefore, we use the variables $(\rho, \rho v, E, q)$ with $q=s^{\textrm{(neq)}}_{w}(w;\varepsilon)$ and rewrite the balance laws \eqref{eq:CDF-w} as
\begin{equation}\label{eq:CDF-q}
	\begin{aligned}
		\partial_t \rho + \partial_x (\rho v) &= 0,\\
		\partial_t (\rho v) + \partial_x(\rho v^2 + \rho \theta ) &= 0,\\
		\partial_t E+ \partial_x(E v + q + \rho \theta v) &= 0,\\
		\partial_t q + v \partial_x q + \frac{g}{\rho} \partial_x \theta^{-1} &= \frac{gMq}{\rho}
	\end{aligned}
\end{equation}
with 
$$
g=s^{\textrm{(neq)}}_{ww}(w;\varepsilon)<0.
$$ 
For smooth solutions, the system \eqref{eq:CDF-q} is equivalent to the balance laws \eqref{eq:CDF-w}.
Recall that the non-equilibrium variable $w$ can be globally expressed in terms of $q$ and $\varepsilon$ due to the strict concavity of $s^{\textrm{(neq)}}(w ;\varepsilon)$ with respect to $w$. Then $g$ can be viewed as a function of $(q;\varepsilon)$. Therefore, our task becomes to learn the negative $g=g(q;\varepsilon)$ and the positive $M=M(\rho, e, q;\varepsilon)$.

Because the training data are known only at discrete space-time points, it is a common practice to replace or approximate the PDEs with their discrete versions \cite{han2019uniformly}. 
Here we discretize the last equation in \eqref{eq:CDF-q} as 
\begin{equation*}
	q_j^{n+1} = q_j^{n} - \frac{\Delta t}{2\Delta x}v_j^n( q_{j+1}^n - q_{j-1}^n) - \frac{\Delta t}{2\Delta x} \frac{g_{j}^n}{\rho_j^n} \brac{(\theta_{j+1}^n)^{-1} - (\theta_{j-1}^n)^{-1}} + \Delta t \brac{\frac{gMq}{\rho}}_{j}^n,
\end{equation*}
where the indices $j$ and $n$ together denote the space-time point $(n\Delta t, j\Delta x)$. 
By writing the last equation in the abstract form
\begin{equation}\label{CDFeqs:LFscheme}
	q_j^{n+1} = \mathcal{S}[g,M](V_{j-1}^n,V_j^n,V_{j+1}^n;\Delta t,\Delta x)
\end{equation}
with $V=(\rho,v,E,q)$, we define our loss function as the mean squared error (MSE):
\begin{equation}\label{lossfunction}
	\mathcal{L} = \sum_{\text{training data}} |q_j^{n+1} - \mathcal{S}[g,M](V_{j-1}^n,V_j^n,V_{j+1}^n;\Delta t,\Delta x)|^2.
\end{equation}
Note that the first three equations in \eqref{eq:CDF-q} do not involve $g$ and $M$ and thereby they are irrelevant at this point.
In addition, the equation \eqref{CDFeqs:LFscheme} is just a discrete version of the fourth equation in \eqref{eq:CDF-q}. It should not be understood as a numerical scheme thereof. 

With the loss function defined in \eqref{lossfunction}, 
we use fully-connected neural networks to approximate $g$ and $M$ in \eqref{eq:CDF-q}. 
To ensure the positivity of $M$ and $-g$, the softplus function is added in the output layer. 
% {Notice that the Knudsen number $\varepsilon$ could vary in a wide range ($10^{-3}$ to $10$). By  resorting to Maxwell iteration (\textbf{cite any reference?}), we find that $M$ should be of order $1/\varepsilon$ such that the equations \eqref{eq:CDF-q} tend to Navier-Stokes equations (\textbf{exactly the same with NS equation?}) as $\varepsilon$ goes to zero, which is highly expected. In order to learn satisfactory $M=M(\rho,e,q;\varepsilon)$, we design the network of $M$ by dividing $\varepsilon$ in the output layer. Equivalently, we actually learn a function $\tilde{M}=\tilde{M}(\rho,e,q)$ (\textbf{$\tilde{M}$ depend on $\varepsilon$ or not?}) and take $M(\rho,e,q;\varepsilon)=\tilde{M}/\varepsilon$.}
Moreover, only smooth solutions are taken as training data since it is not clear how to define weak solutions for the non-conservative system \eqref{eq:CDF-q} \cite{Dafermos2005}.

After having $g$ and $M$, we need another neural network to get the balance laws \eqref{eq:CDF-w} where $s^{\textrm{(neq)}}_w=s^{\textrm{(neq)}}_w(w;\varepsilon)$ is to be determined. To do this, we recall from $q=s^{\textrm{(neq)}}_{w}(w;\varepsilon)$ that 
$w$ can be solved as a function $w=F(q;\varepsilon)$ satisfying
$$
q=s^{\textrm{(neq)}}_{w}(F(q;\varepsilon);\varepsilon). 
$$
From this we have the relation
\begin{equation*} 
	g=s^{\textrm{(neq)}}_{ww}(w;\varepsilon)=s^{\textrm{(neq)}}_{ww}(F(q,\varepsilon);\varepsilon)=\frac{1}{ F_q(q;\varepsilon)}.
\end{equation*}
Namely, $F_q(q;\varepsilon)$ has been trained. 

Then we design a fully-connected neural network $\mathcal{F}(q;\varepsilon)$ to approximate $F(q;\varepsilon)$ which should be strictly decreasing. 
With the automatic differentiation technique \cite{baydin2017automatic} to compute the derivative $\mathcal{F}_q(q;\varepsilon)$ with respect to $q$, we define the loss function as
\begin{equation}\label{loss of F}
	\mathcal{L}_F = \sum_{q} \left|\frac{1}{g(q;\varepsilon)} - \mathcal{F}_q(q;\varepsilon)\right|^2.
\end{equation}
Here the summation is over some discrete $q$ in a reasonable interval.  
Having $F(q;\varepsilon)$, the function $s^{\textrm{(neq)}}_{w}(w;\varepsilon)$ can be uniquely determined by solving the nonlinear algebraic equation $w=F(q;\varepsilon)$ with the  bisection method. In this way, the balance laws \eqref{eq:CDF-w} are fixed.

We conclude this section with the following remarks.

\begin{rem}
The learned balance laws \eqref{eq:CDF-w} can be used to predict non-smooth solutions although they are learned through the equivalent version \eqref{eq:CDF-q} only with smooth training data.
In the prediction, we will use the Lax-Friedrichs scheme which is irrelevant to \eqref{CDFeqs:LFscheme}. 
\end{rem}
\begin{rem}
	Instead of resorting to the non-conservative form \eqref{eq:CDF-q}, we have also explored the possibility in learning the balance laws \eqref{eq:CDF-w} directly. However, the computational cost is much larger. The reason is that the algebraic equation $q=s^{\textrm{(neq)}}_{w}(w;\varepsilon)$ has to be solved to obtain $w$ whenever the neural network defining the equation is updated. 
\end{rem}

\begin{rem}
In learning the function $F(q;\varepsilon)$, we have also tried to use the convex neural network \cite{amos2017input} to enforce its monotonicity with respect to $q$. But the performance does not seem to be better. 
\end{rem}

\section{Numerical results}\label{sec:numerical}

This section contains three subsections, where we present some results of the learning procedure given in the previous section.
The domain of the balance laws \eqref{eq:CDF-w} is taken to be $(x,t)\in [-\pi, \pi]\times [0,0.5]$ with periodic boundary conditions.

\subsection{Generating Data}

In this part, we show how to generate the training data. 
% describe the details on the generation of data sets.
To do this, we solve the BGK model \eqref{eq:kinetic-bgk} in the domain  $(x,t)\in [-\pi, \pi]\times [0,0.5]$ with periodic boundary conditions and initial data constructed below. Once the distribution $f=f(x,t;\xi)$ is solved, the training data $(\rho,\rho v,E,q)$ can be obtained by using \eqref{rhovT} and \eqref{E and q}.

For the purpose of computation, the velocity space $\xi\in \mathbb{R}$ is truncated into a sufficiently large interval $\xi\in[-\abs{\xi}_{\max}, \abs{\xi}_{\max}]$, which is discretized into $N_\xi$ grid points with $\Delta \xi=2\abs{\xi}_{\max}/N_\xi$. 
The spatial interval $[-\pi, \pi]$ is discretized into $N_x$ grid points with $\Delta x=2\pi/N_x$. To ensure that the numerical error is sufficiently small, we use high-order numerical methods: the fifth-order WENO method \cite{jiang1996efficient} in the space variable and the third-order implicit-explicit (IMEX) method \cite{ascher1997implicit} in time. 

The time step is taken to be $\Delta t=0.1\Delta x$. We fix the parameters $\abs{\xi}_{\max}=10$ and $N_{\xi}=100$ such that the discretization error in velocity is much smaller than that in space and time. The number of grid points in space is taken to be $N_x=80$. Our numerical experience shows that the numerical error is generally less than $10^{-6}$ with these parameters for initial data below.

Following \cite{han2019uniformly}, we consider two types of initial conditions. The first one is the smooth function $f_{\textrm{smooth}}$. It is a convex combination of two Maxwellians $f_M(\xi; U_1)$ and $f_M(\xi; U_2)$:
\begin{equation*}
	f_{\textrm{smooth}} = \alpha f_M(\xi; U_1) + (1-\alpha) f_M(\xi; U_2)
\end{equation*}
with $\alpha$ sampled from $[0,1]$.
Here the macroscopic variables $U_i=(\rho_i, v_i, T_i)$ for $i=1,2$ are taken to be the sine waves
\begin{equation}\label{eq:init-sin}
\left\{
\begin{aligned}
\rho_i(x,0) &= a_{\rho,i}\sin({k_{\rho,i}} x+\psi_{\rho,i}) + b_{\rho,i},  \\
v_i(x,0) &= 0, \\
T_i(x,0) &= a_{T,i}\sin({k_{T, i}} x+\psi_{T,i}) + b_{T,i}.
\end{aligned}
\right.
\end{equation}
In \eqref{eq:init-sin}, the parameters $a_{\rho,i}$ and $a_{T, i}$ are sampled from $[0.2, 0.3]$, {$k_{\rho,i}$ and $k_{T, i}$ are from $\{1, 2\}$}, $\psi_{\rho,i}$ and $\psi_{T, i}$ are from $[0, 2\pi]$, and $b_{\rho,i}$ and $b_{T, i}$ from $[0.5, 0.7]$.
Note that with such parameters, $\rho_i$ and $T_i$ are always positive.

The second type of initial conditions are of the form
\begin{equation*}
	f_{\textrm{shock}} = \alpha f_M(\xi;U_{\textrm{smooth}}) + (1-\alpha) f_M(\xi;U_{\textrm{shock}})
\end{equation*}
with $\alpha$ sampled from $[0, 1]$. 
Here $U_{\textrm{smooth}}$ is taken to be $U_1$ or $U_2$ above, 
and 
\begin{equation}\label{5.4}
U_{\textrm{shock}}(x,0) = 
\left\{
\begin{aligned}
& (\rho_1, 0, T_1), \quad x\in[-\pi, x_1]\cup[x_2,\pi], \\
& (\rho_2, 0, T_2), \quad x\in(x_1,x_2). \\
\end{aligned}
\right.
\end{equation}
In \eqref{5.4}, $\rho_1$ and $T_1$ are two constants sampled from $[1, 1.1]$; $\rho_2$ and $T_2$ from $[0.55, 0.65]$; and $x_1$ and $x_2$ are the discontinuity locations sampled from $[-2, -1.8]$ and $[1.5, 1.7]$, respectively.

\subsection{Training}

To train the functions $g=g(q;\varepsilon)$ and $M=M(\rho, e, q;\varepsilon)$ in \eqref{eq:CDF-q}, we design two fully-connected neural networks based on the loss function \eqref{lossfunction}. 
Each of them has three hidden layers with 30 nodes in each layer. 
As the activation function, the softplus function is used in the output layer to ensure the positivity of $M$ and $-g$, while the $\tanh$ function is used in other layers. 
% {In the output layer of $M$, we divided by Knudsen  number $\varepsilon$.}
The networks are trained by using back propagation with the stochastic gradient descent (SGD) algorithm \cite{ruder2016overview}. The learning rate is taken to be 0.05, the momentum is 0.9, and the batch size is 50.

In our training, we only use $50$ different initial values of the first type \eqref{eq:init-sin} with $k_{\rho,i}=k_{T,i}=1$. The training data are the solutions, initiated at these initial values, evaluated at the space-time points $(x_j, t^n)=(-\pi + j \Delta x, n\Delta t)$ with $j=1, 2, \cdots, 80$ and $n=0, 1, \cdots, 10$. Here and in the loss function \eqref{lossfunction} $\Delta t=0.1$ and $\Delta x= 2\pi/80$. Note that only the smooth initial data $f_{\textrm{smooth}}$ (the first type) are used.  

To train the inverse function $w=F(q;\varepsilon)$ of $q=s^{\textrm{(neq)}}_{w}(w;\varepsilon)$ in \eqref{eq:CDF-w}, we use the loss function \eqref{loss of F} and design another fully-connected neural network $\mathcal{F}(q;\varepsilon)$ of three hidden layers with 20 nodes in each layer. 
For computing the derivative $\mathcal{F}_q(q;\varepsilon)$ in \eqref{loss of F}, the automatic differentiation technique \cite{baydin2017automatic} is employed.
In addition, we approximate $w=F(q;\varepsilon)$ with $\mathcal{F}(q;\varepsilon)-\mathcal{F}(0;\varepsilon)$, which ensures $F(0;\varepsilon)=0$ and thereby the uniqueness. 

\begin{figure} 
    \centering
    \includegraphics[width=0.8\textwidth]{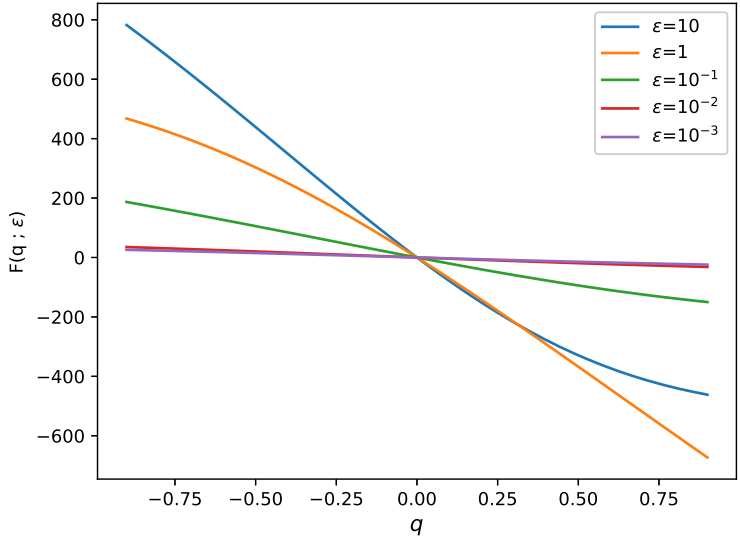}
    \caption{Profiles of function $w=F(q;\varepsilon)$ for different Knudsen numbers $\varepsilon=10^{-3}, 10^{-2}, 10^{-1}, 1, 10$. The intersection point reflects $F(0;\varepsilon)=0$.}
    \label{fig:f-profile}
\end{figure}

The learned $w=F(q;\varepsilon)$ is illustrated in Figure \ref{fig:f-profile}. Clearly, it is a strictly decreasing function of $q$ for each Knudsen number $\varepsilon$. This monotonicity guarantees the strictly concavity of the entropy function $s^{\textrm{(neq)}}=s^{\textrm{(neq)}}(w;\varepsilon)$ and thereby the conservation-dissipation principle.

\subsection{Prediction}

To show the performance of the CDF-based machine learning model \eqref{eq:CDF-w}, we solve it numerically and compare the numerical solutions with those computed from the kinetic equation \eqref{eq:kinetic-bgk}. 
To compute the solutions, we discretize the learned balance laws \eqref{eq:CDF-w} by using the Lax-Friedrichs scheme with a relatively fine mesh. The scheme is first-order, monotone and can produce entropy solutions in the presence of shocks. 

Let $\hat{U}=(\hat{\rho}, \hat{\rho}\hat{v}, \hat{E})$ be the solution computed by the scheme and $U=(\rho, \rho v, E)$ be the solution generated by the BGK model \eqref{eq:kinetic-bgk} with the same initial data. 
We define their relative $L^1$ and $L^2$ errors as 
\begin{equation*}
	E_1 := \frac{\sum_{j}\abs{U_j - \hat{U}_j}}{\sum_{j}\abs{U_j}},
\end{equation*}
and
\begin{equation*}
	E_2 := \sqrt{\frac{\sum_{j}(U_j - \hat{U}_j)^2}{\sum_{j}(U_j)^2}},
\end{equation*}
where $U_j$ and $\hat{U}_j$ are the values of $U$ and $\hat{U}$ at the $j$-th spatial grid point $x_j$ and the final time $t=0.5$. {We note that the prediction solution $\hat{U}$ is computed by the Lax-Friedrich   scheme with a refined mesh grid $N_x=400$.}

The relative errors for smooth and discontinuous solutions are given in Table \ref{tab:error-smooth} and Table \ref{tab:error-discontinuous}, respectively. 
The mean and the standard derivation (std) are taken from 10 randomly different runs. From Table \ref{tab:error-smooth} and Table \ref{tab:error-discontinuous}, we see that the magnitude of relative errors is less than 3\% for smooth solutions and 6\% for discontinuous solutions. 
Moreover, the errors become larger for larger Knudsen number $\varepsilon$, which is reasonable 
because it is more difficult to model non-equilibrium flows. 

It is remarkable that although only the smooth training data are used to learn the balance laws \eqref{eq:CDF-w}, the learned model performs well also with the discontinuous initial data.

\begin{table}[htbp]
  \centering
    \begin{tabular}{c|c|c|c|c}
    \hline
    \multirow{2}[4]{*}{} & \multicolumn{2}{c|}{relative $L^1$ error} & \multicolumn{2}{c}{relative $L^2$ error} \bigstrut\\
\cline{2-5}  Knudsen number $\varepsilon$  &   mean     &   std         &  mean     &   std    \bigstrut\\
    \hline
    $10^{-3}$ 	&   2.46e-03 & 4.13e-04 & 2.62e-03 & 4.32e-04  \bigstrut[t]\\
    $10^{-2}$ 	&   1.72e-03 & 1.57e-04 & 1.89e-03 & 1.35e-04  \\
    $10^{-1}$ 	&   7.83e-03 & 2.62e-03 & 8.39e-03 & 2.71e-03  \\
    $10^0$ 		&   2.17e-02 & 6.68e-03 & 2.33e-02 & 7.14e-03  \\
    $10^1$ 		&   2.18e-02 & 5.37e-03 & 2.35e-02 & 5.73e-03  \bigstrut[b]\\
    \hline
    \end{tabular}%
     \caption{Relative $L^1$ and $L^2$ errors for smooth solutions at $t=0.5$. The mean and the standard derivation (std) are taken from 10 randomly different runs.}
  \label{tab:error-smooth}%
\end{table}%

\begin{table}[htbp]
  \centering
    \begin{tabular}{c|c|c|c|c}
    \hline
    \multirow{2}[4]{*}{} & \multicolumn{2}{c|}{relative $L^1$ error} & \multicolumn{2}{c}{relative $L^2$ error} \bigstrut\\
\cline{2-5}  Knudsen number $\varepsilon$  &   mean     &   std      &  mean     &   std     \bigstrut\\
    \hline
    $10^{-3}$ 	& 1.03e-02 & 1.12e-03 & 1.49e-02 & 2.24e-03 \bigstrut[t]\\
    $10^{-2}$ 	& 6.30e-03 & 1.06e-03 & 8.60e-03 & 1.81e-03  \\
    $10^{-1}$ 	& 2.25e-02 & 2.65e-03 & 2.66e-02 & 3.02e-03  \\
    $10^0$ 		& 4.54e-02 & 6.32e-03 & 5.16e-02 & 6.62e-03  \\
    $10^1$ 		& 5.37e-02 & 1.06e-02 & 6.05e-02 & 1.20e-02  \bigstrut[b]\\
    \hline
    \end{tabular}%
      \caption{Relative $L^1$ and $L^2$ errors for discontinuous solutions at $t=0.5$. The mean and the standard derivation (std) are taken from 10 randomly different runs.}
  \label{tab:error-discontinuous}%
\end{table}%

\begin{figure}
\footnotesize
\stackunder[5pt]{\includegraphics[width=0.27\textwidth]{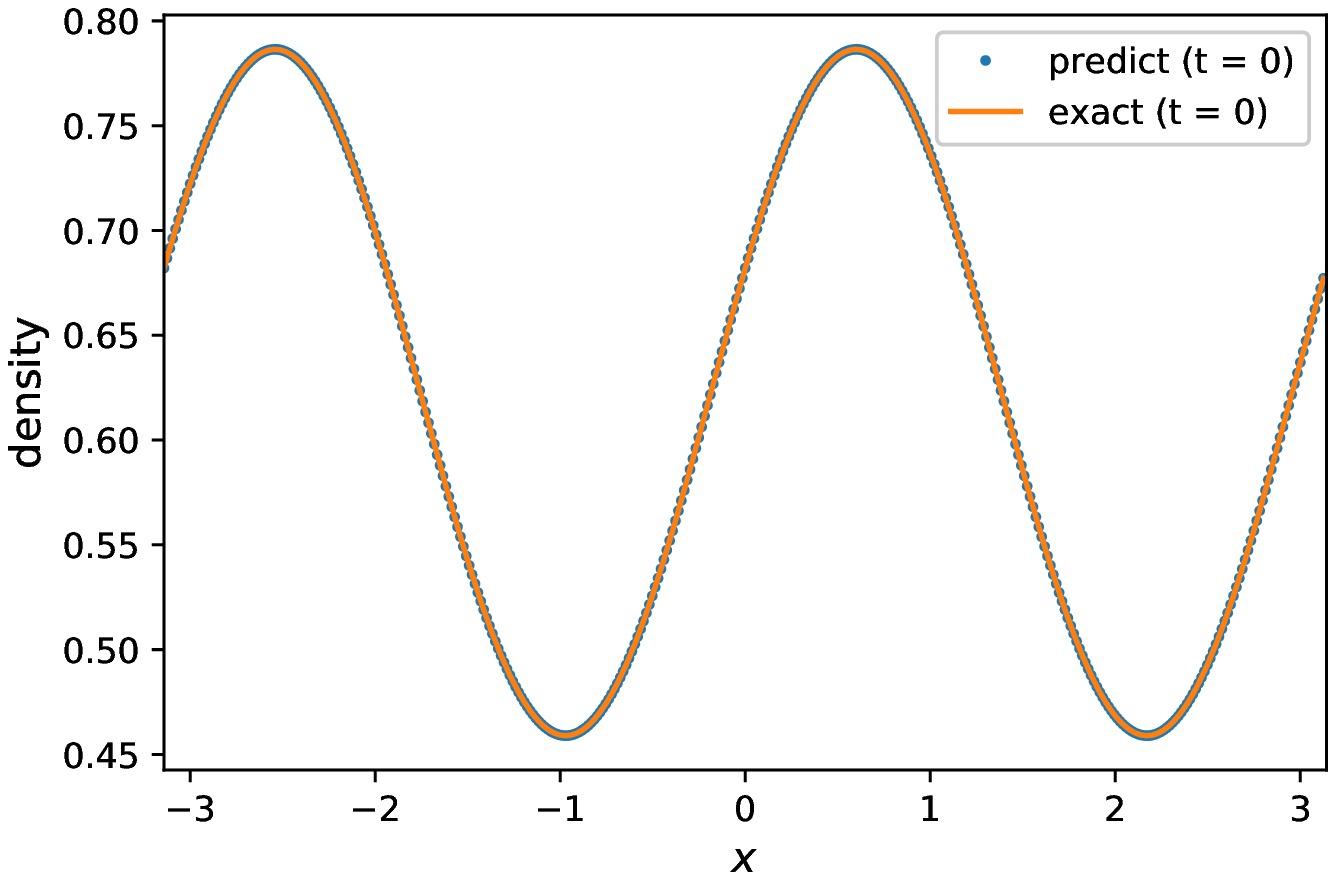}}{(i) density at $t=0$}
\hspace{1cm}%
\stackunder[5pt]{\includegraphics[width=0.27\textwidth]{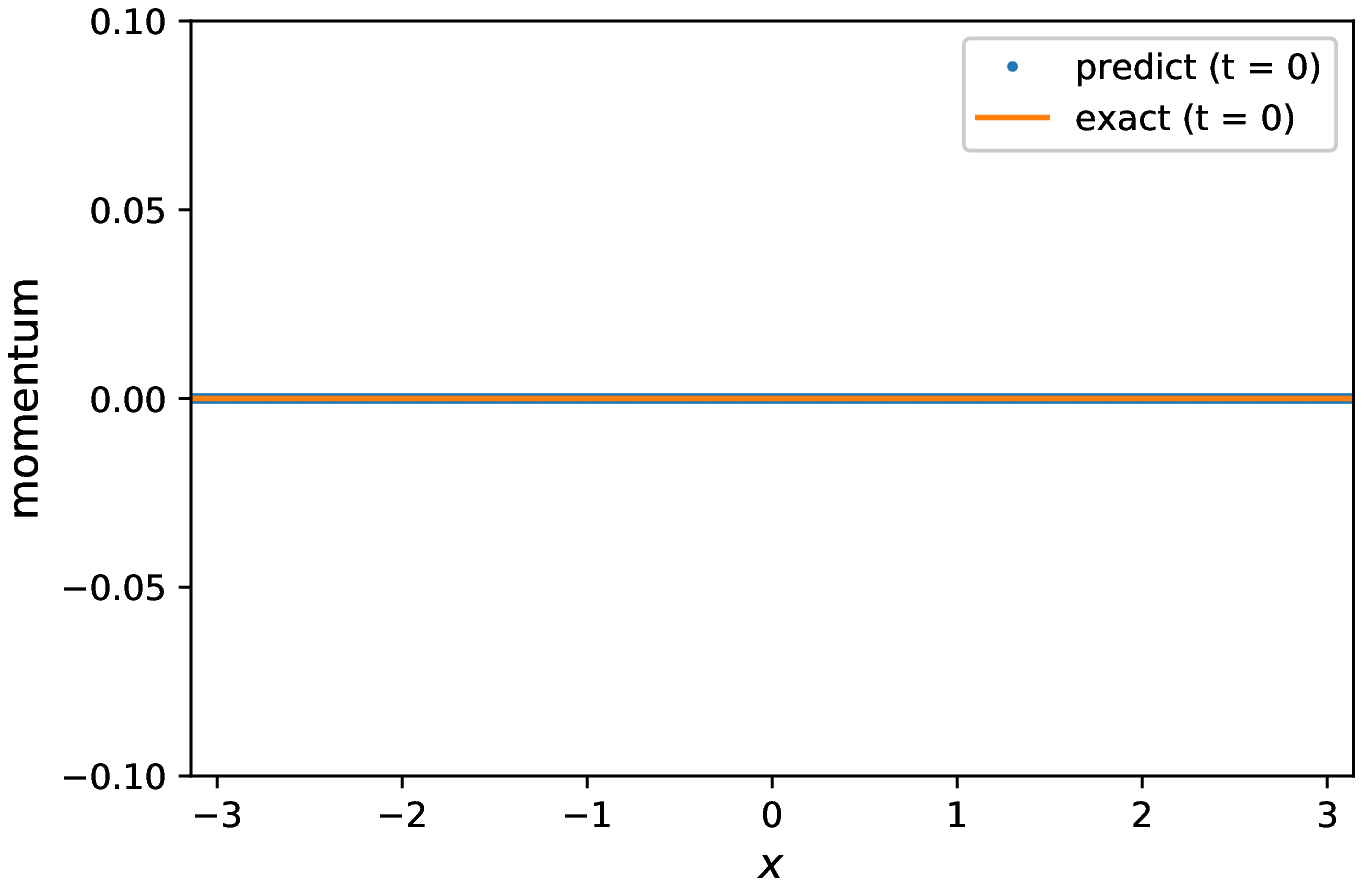}}{(ii) momentum at $t=0$}
\hspace{1cm}%
\stackunder[5pt]{\includegraphics[width=0.27\textwidth]{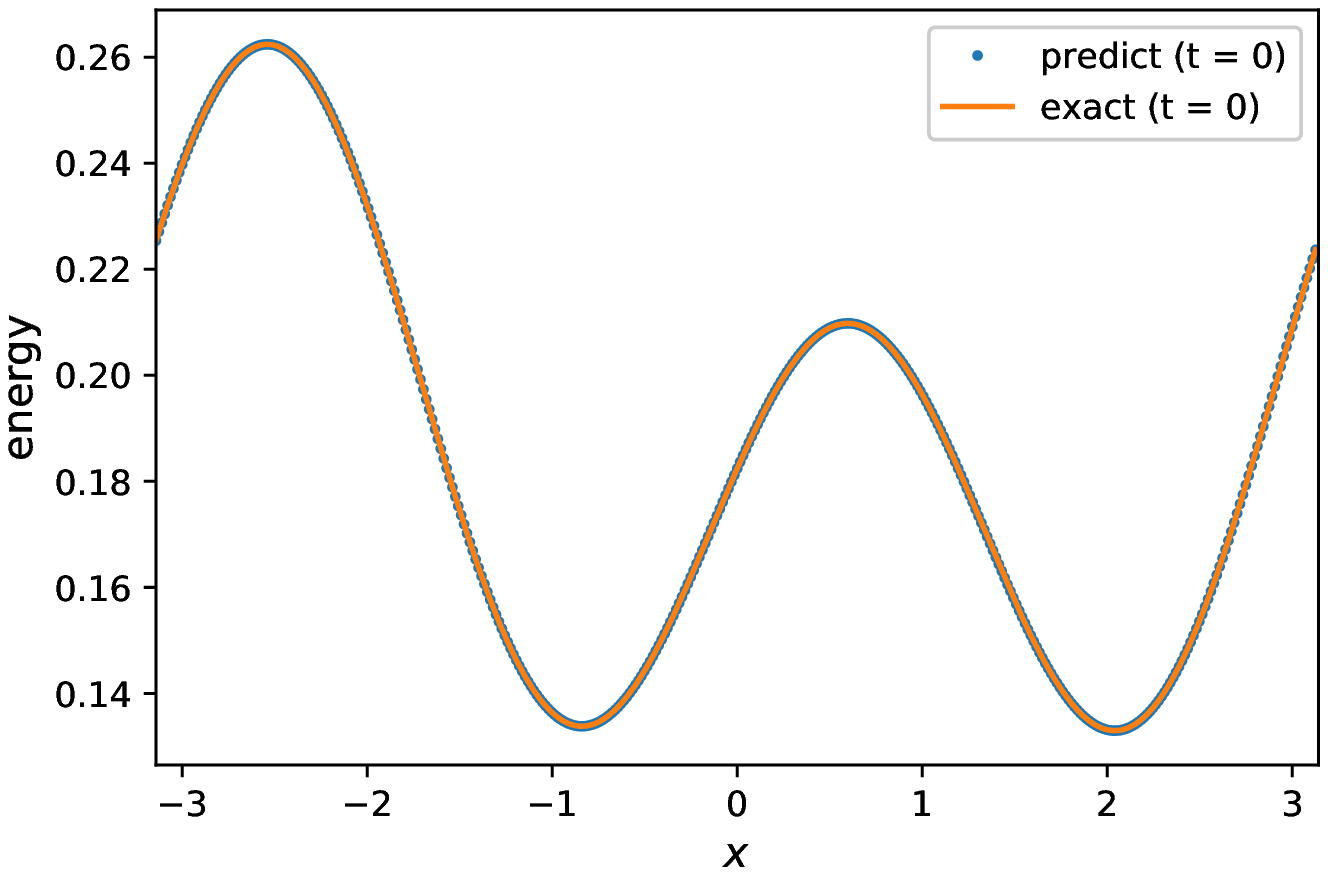}}{(iii) energy at $t=0$}
\bigskip
\stackunder[5pt]{\includegraphics[width=0.27\textwidth]{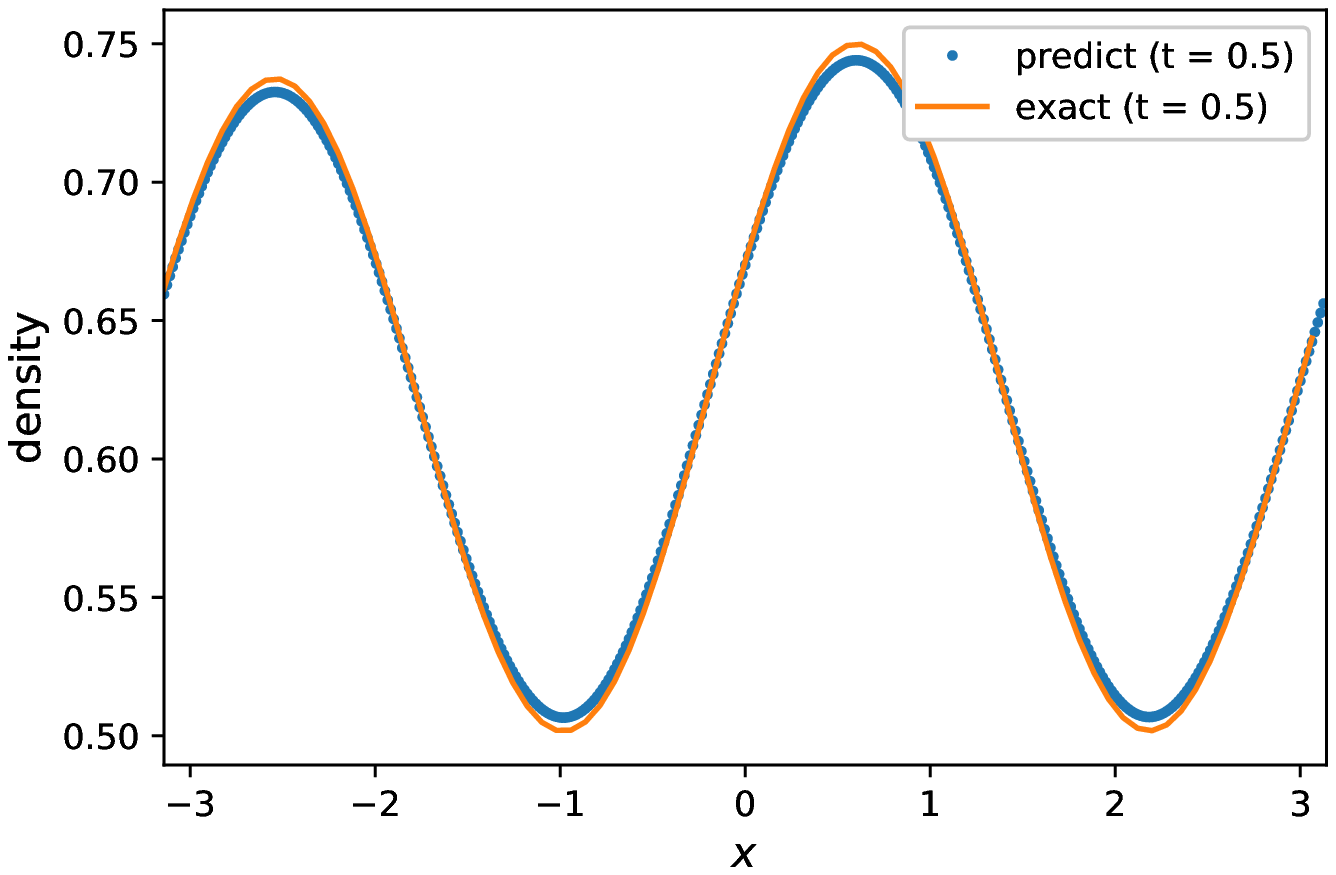}}{(i') density at $t=0.5$}
\hspace{1cm}%
\stackunder[5pt]{\includegraphics[width=0.27\textwidth]{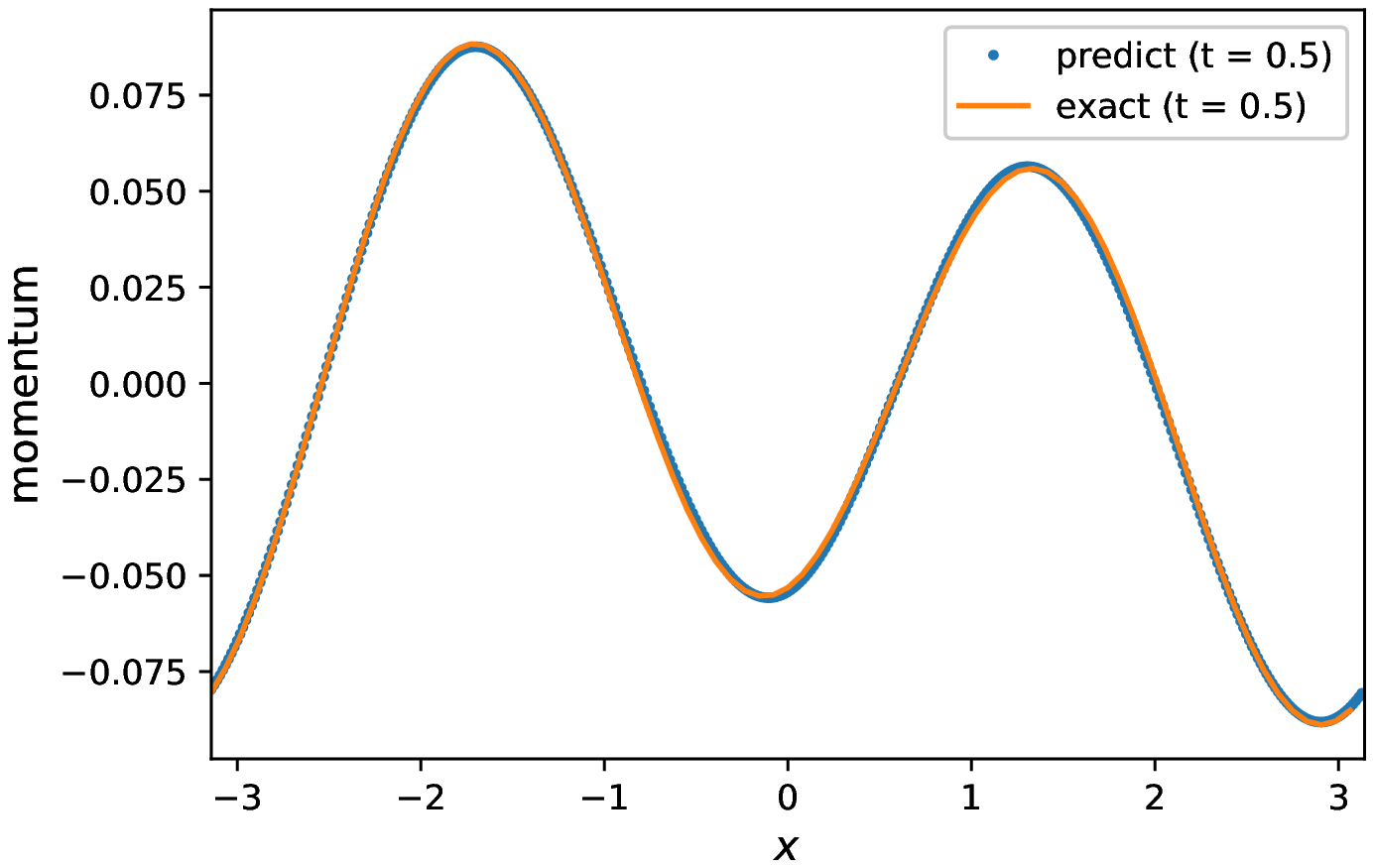}}{(ii') 
momentum at $t=0.5$}
\hspace{1cm}%
\stackunder[5pt]{\includegraphics[width=0.27\textwidth]{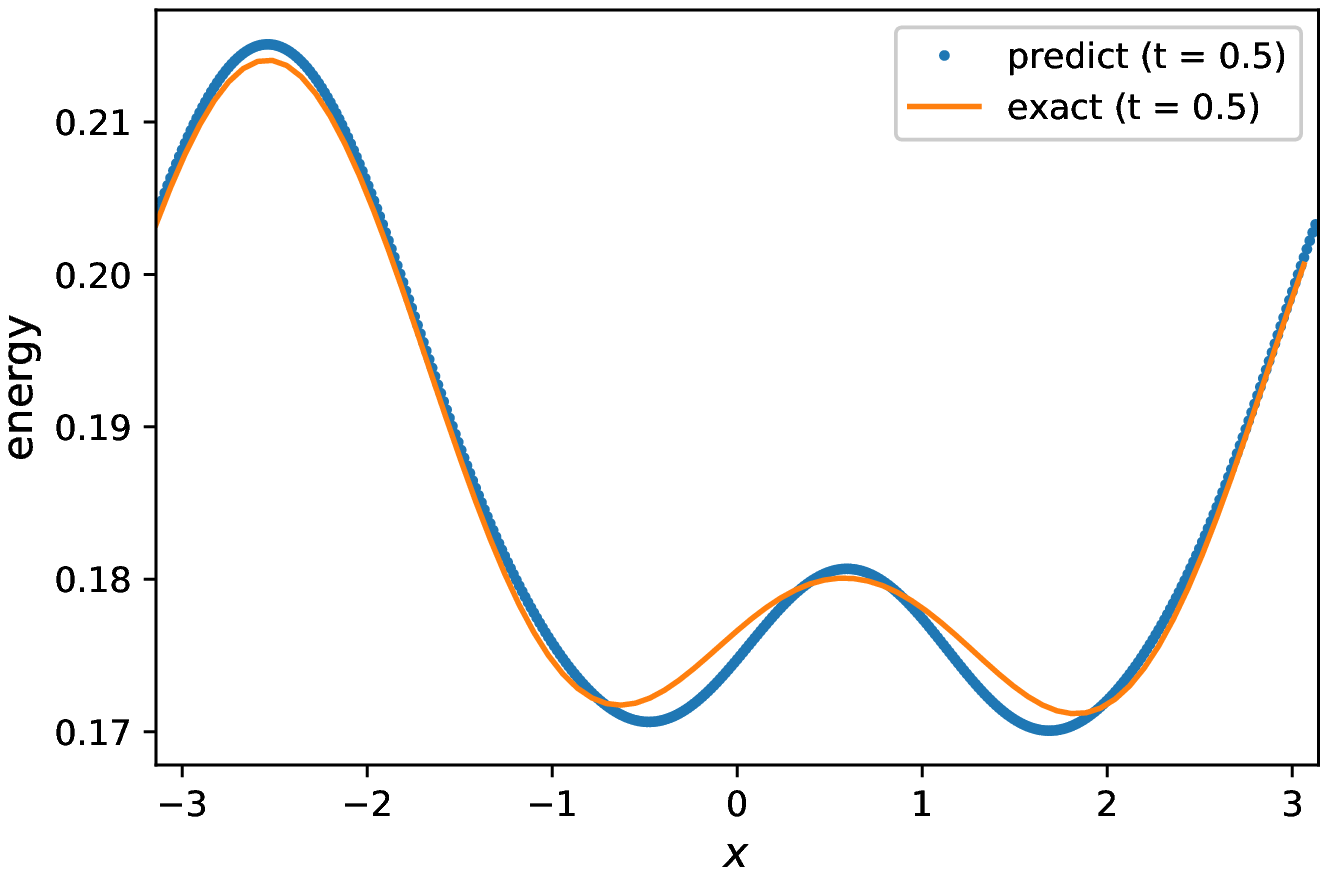}}{(iii') 
energy at $t=0.5$}
\caption{Solution profiles of density, momentum, and energy (from left to right) at $t=0$ and $t=0.5$ (from top to bottom) with $\varepsilon=10$ and with smooth initial data.}
    \label{fig:smooth-Kn1e0}
\end{figure}

Furthermore, we plot the solution profiles of density, momentum, and energy obtained from the kinetic equation and the learned model at $t=0$ and $t=0.5$. 
Figure \ref{fig:smooth-Kn1e0} is for smooth initial data with {$k_{\rho,i}=k_{T,i}=2$ in \eqref{eq:init-sin}, which are not used in the training data,} and with Knudsen number $\varepsilon=10$.
Here and below, the `exact' means that obtained from the kinetic equation while `predict' is for the learned model.
It is observed that the solution profiles of our model agree well with those of the BGK model. The performances with other Knudsen numbers are similar and are omitted here.

\begin{figure}
\footnotesize
\stackunder[5pt]{\includegraphics[width=0.27\textwidth]{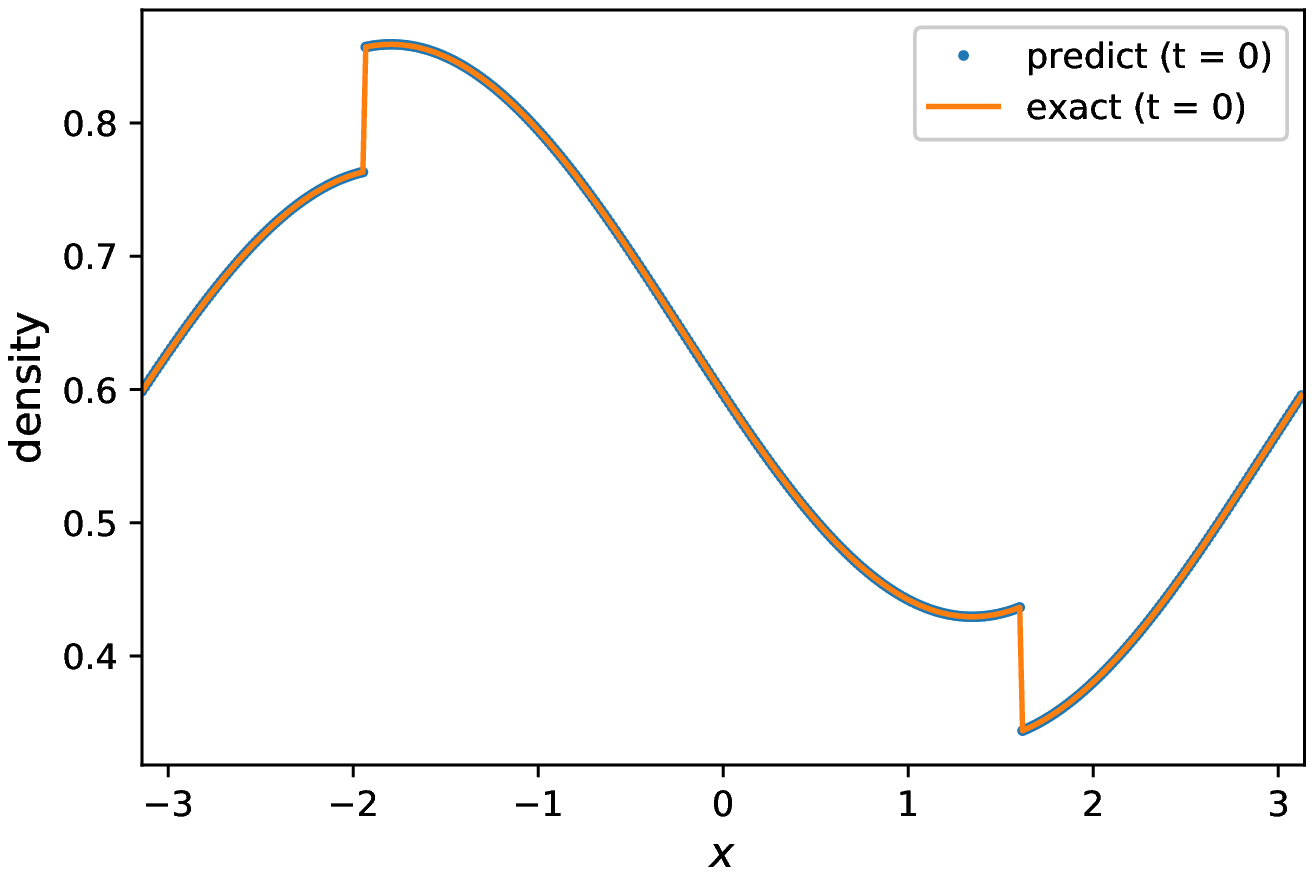}}{(i) density at $t=0$}
\hspace{1cm}%
\stackunder[5pt]{\includegraphics[width=0.27\textwidth]{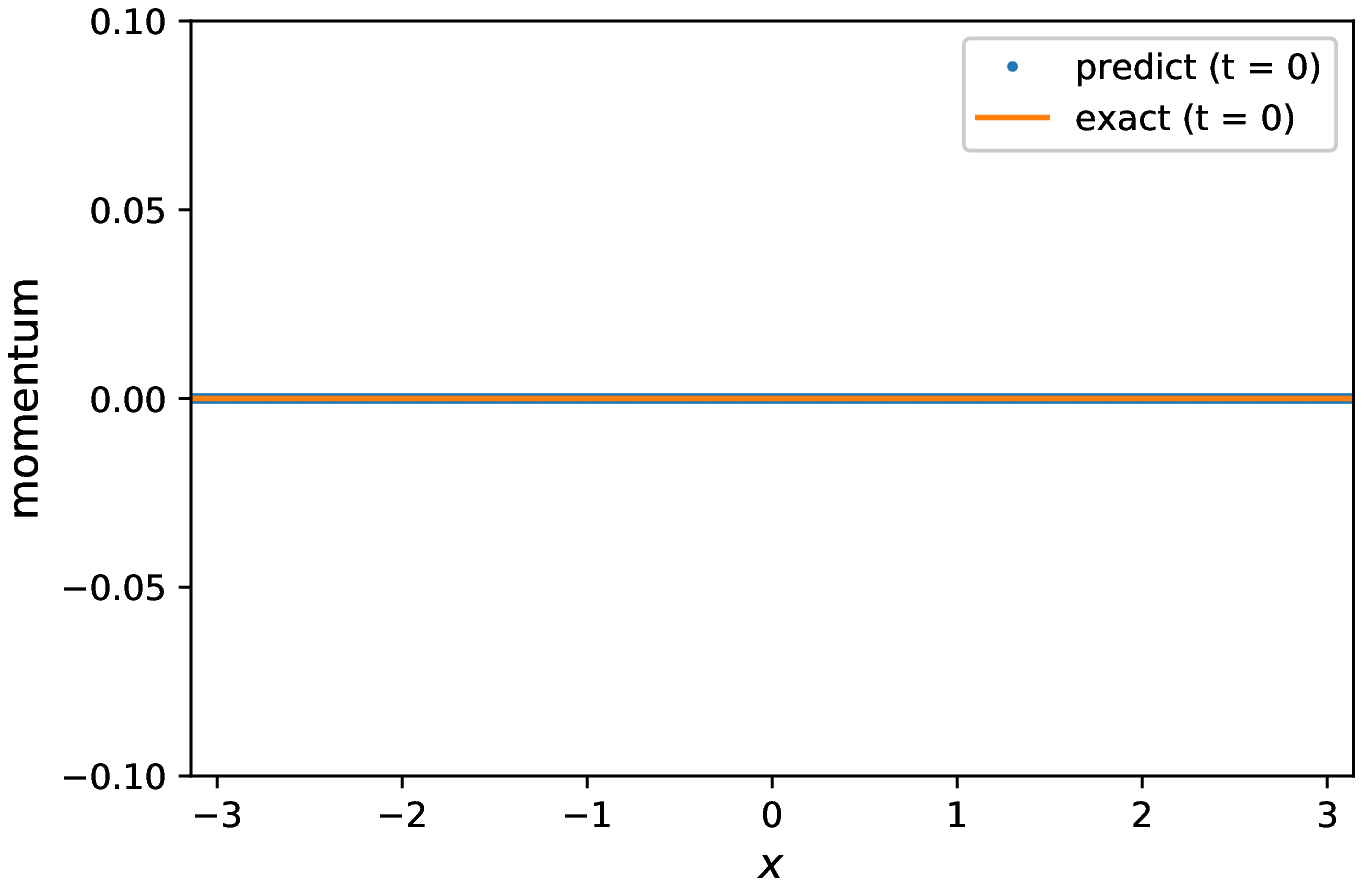}}{(ii) momentum at $t=0$}
\hspace{1cm}%
\stackunder[5pt]{\includegraphics[width=0.27\textwidth]{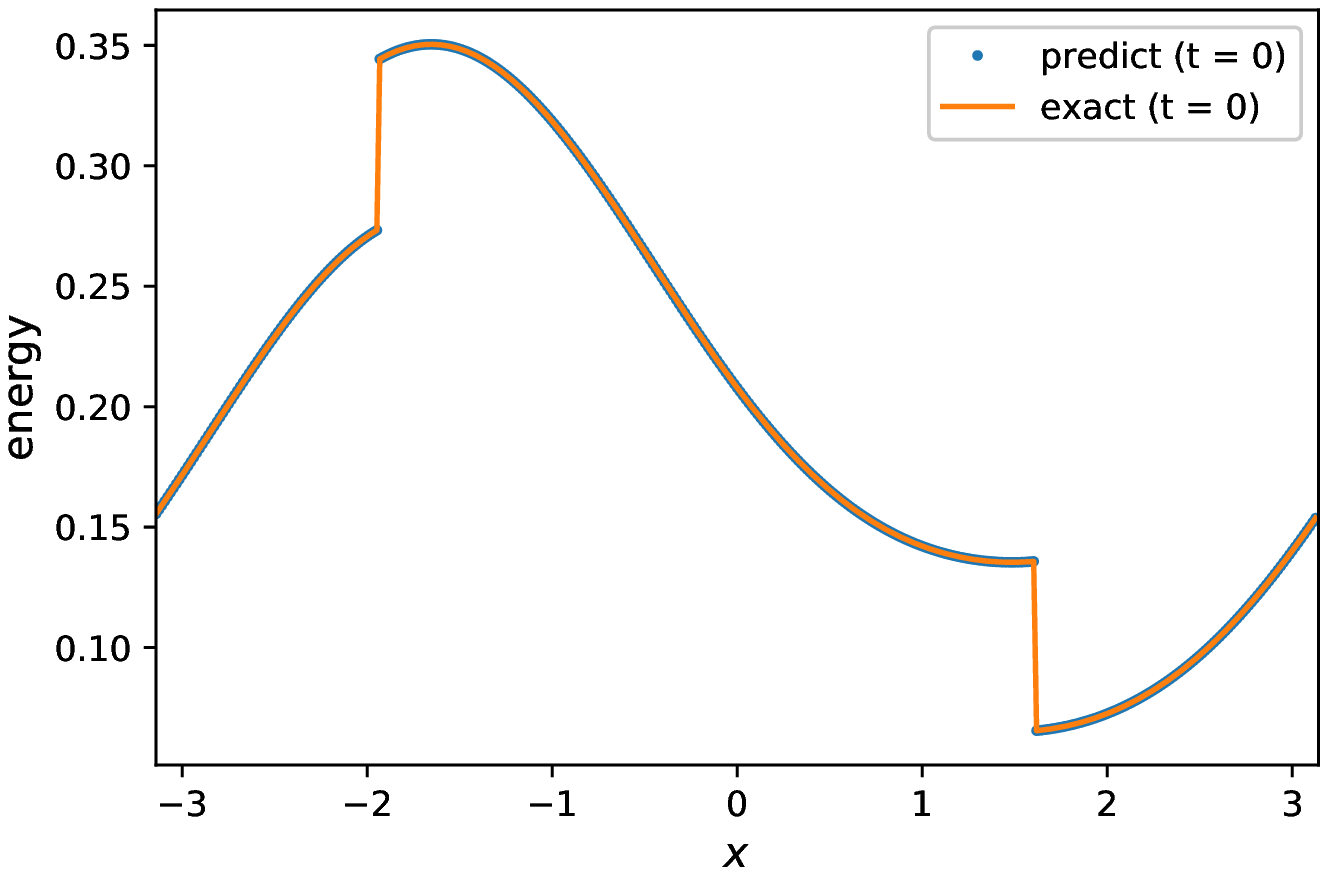}}{(iii) energy at $t=0$}
\bigskip
\stackunder[5pt]{\includegraphics[width=0.27\textwidth]{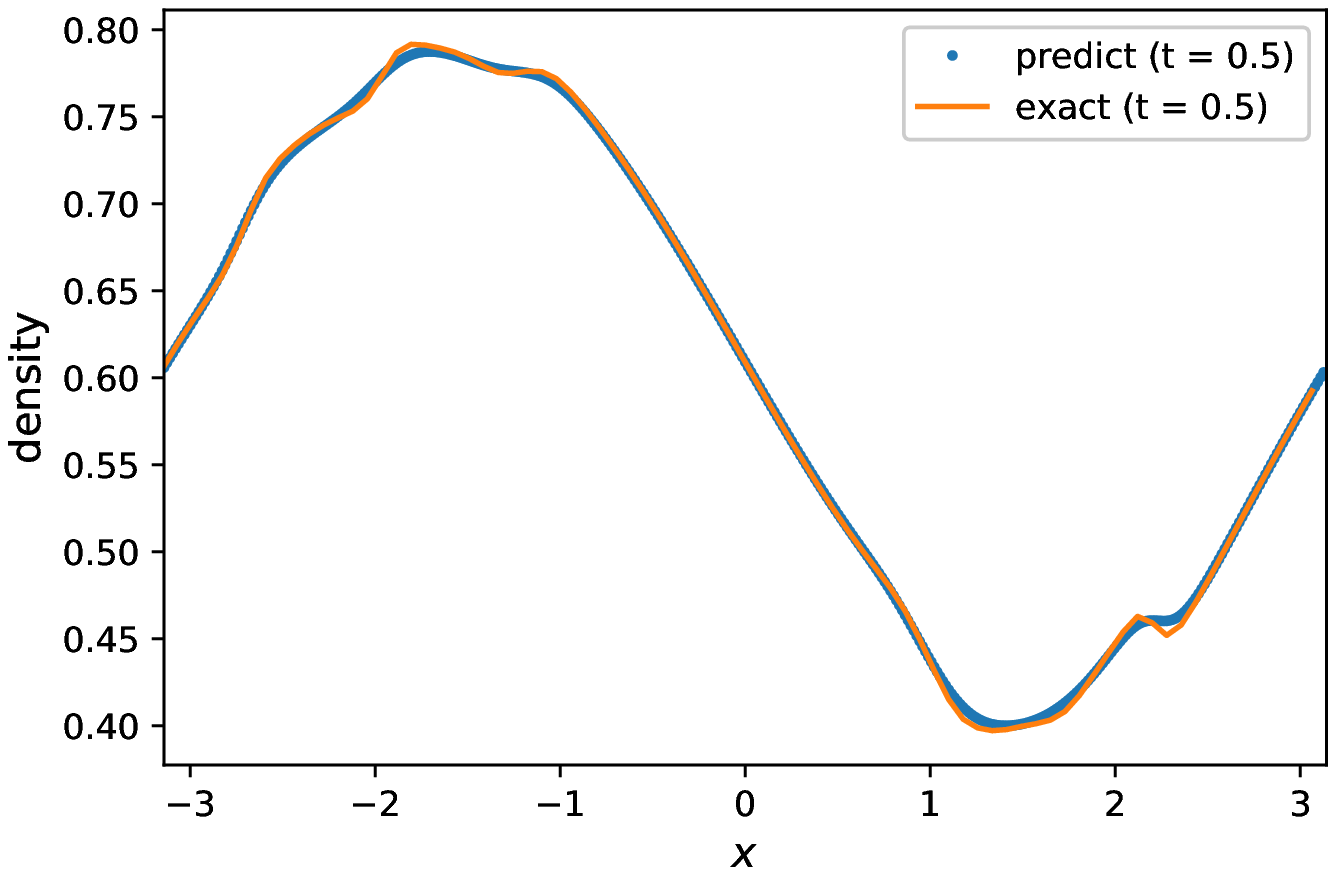}}{(i') density at $t=0.5$}
\hspace{1cm}%
\stackunder[5pt]{\includegraphics[width=0.27\textwidth]{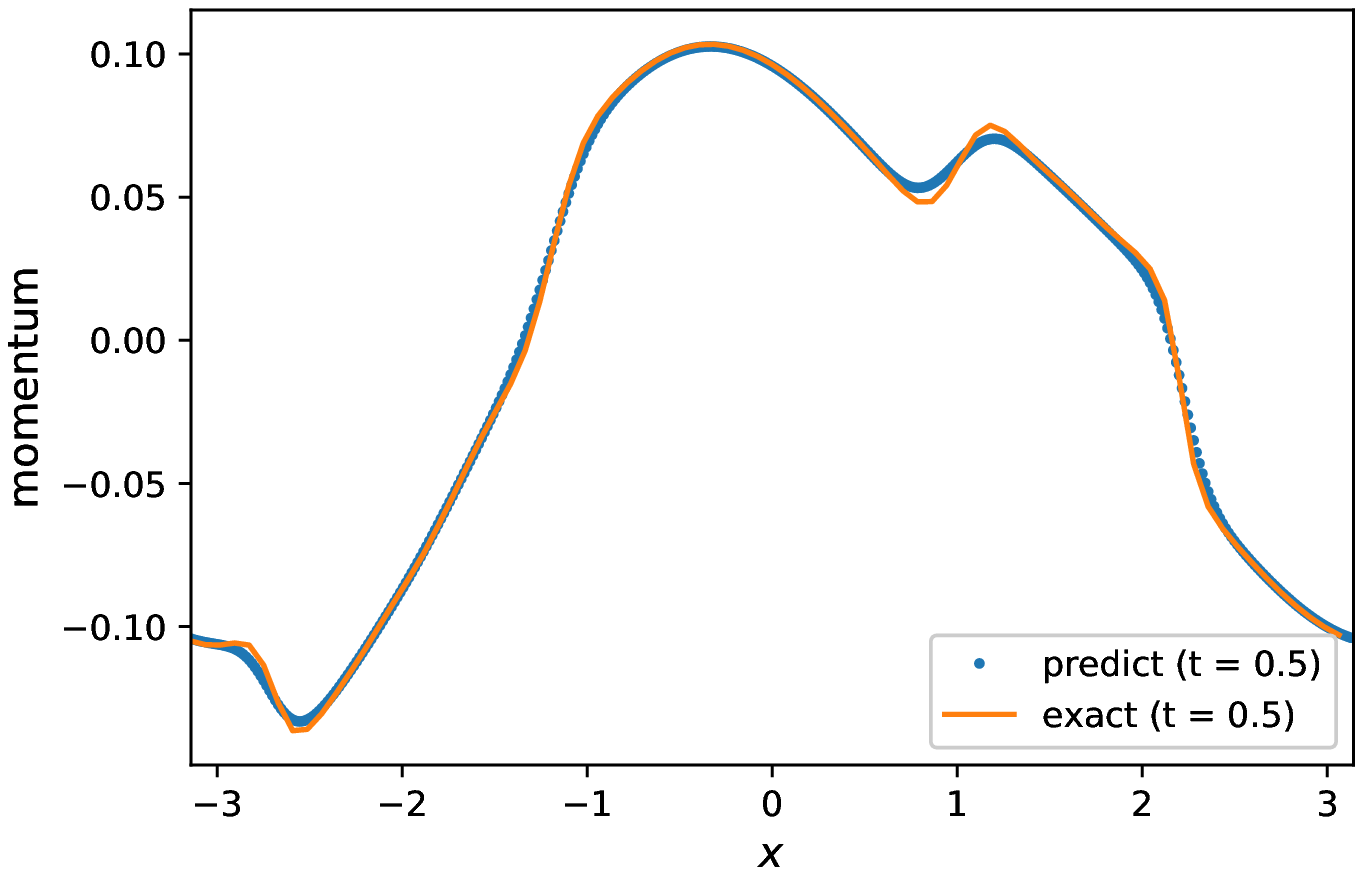}}{(ii') 
momentum at $t=0.5$}
\hspace{1cm}%
\stackunder[5pt]{\includegraphics[width=0.27\textwidth]{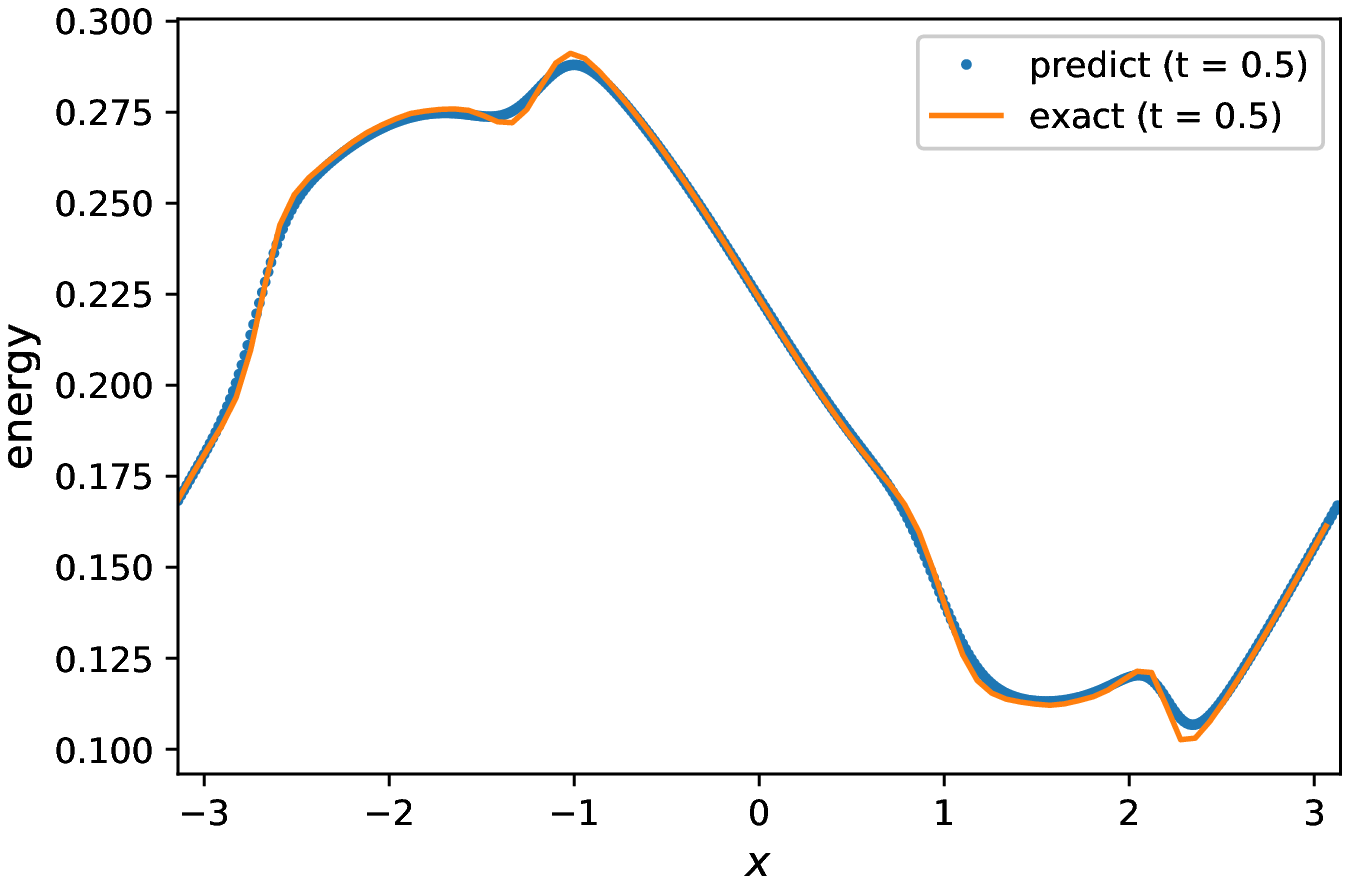}}{(iii') 
energy at $t=0.5$}
\caption{Solution profiles of density, momentum, and energy (from left to right) at $t=0$ and $t=0.5$ (from top to bottom) with $\varepsilon=10^{-3}$ and with discontinuous initial data.}
    \label{fig:shock-Kn1e-3}
\end{figure}

\begin{figure}
\footnotesize
\stackunder[5pt]{\includegraphics[width=0.27\textwidth]{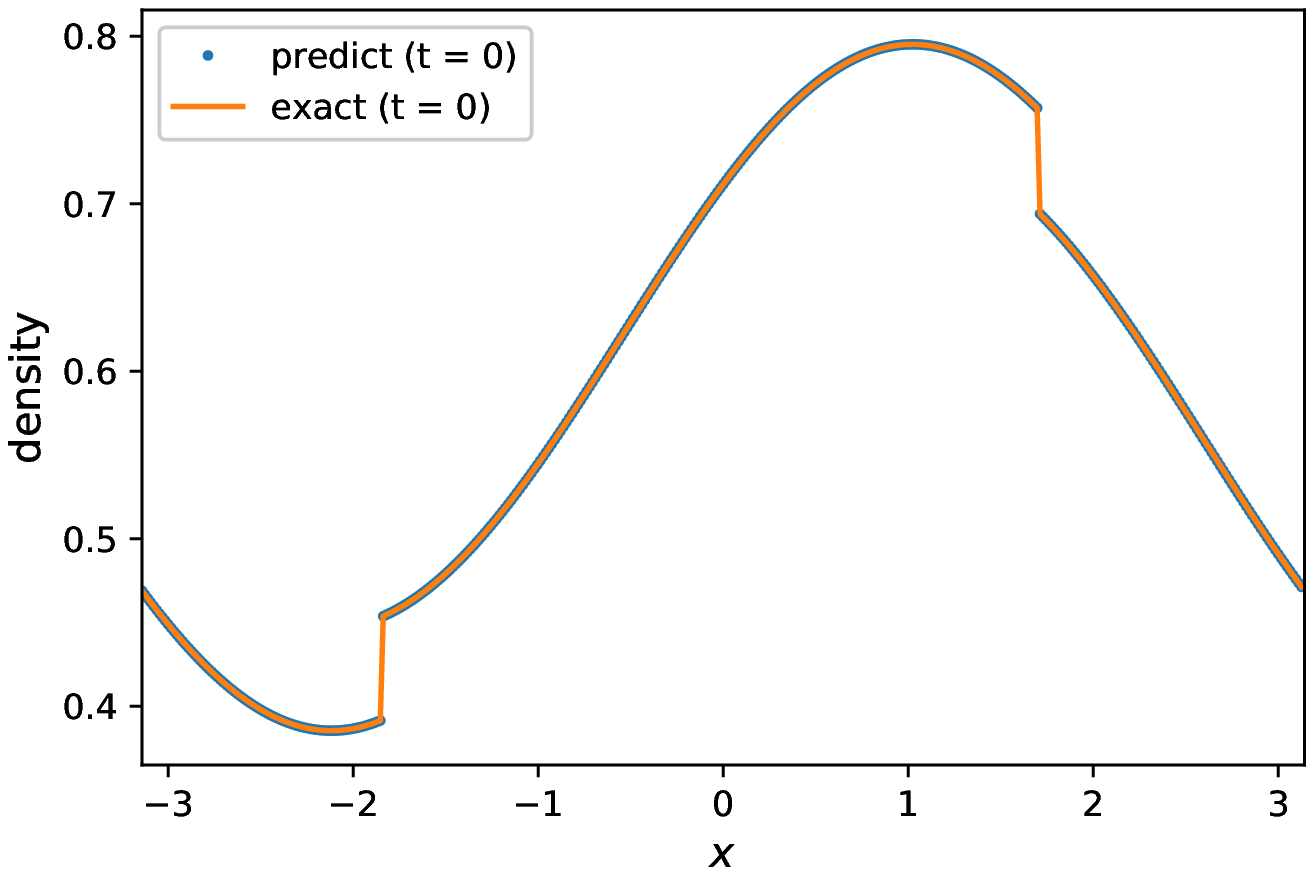}}{(i) density at $t=0$}
\hspace{1cm}%
\stackunder[5pt]{\includegraphics[width=0.27\textwidth]{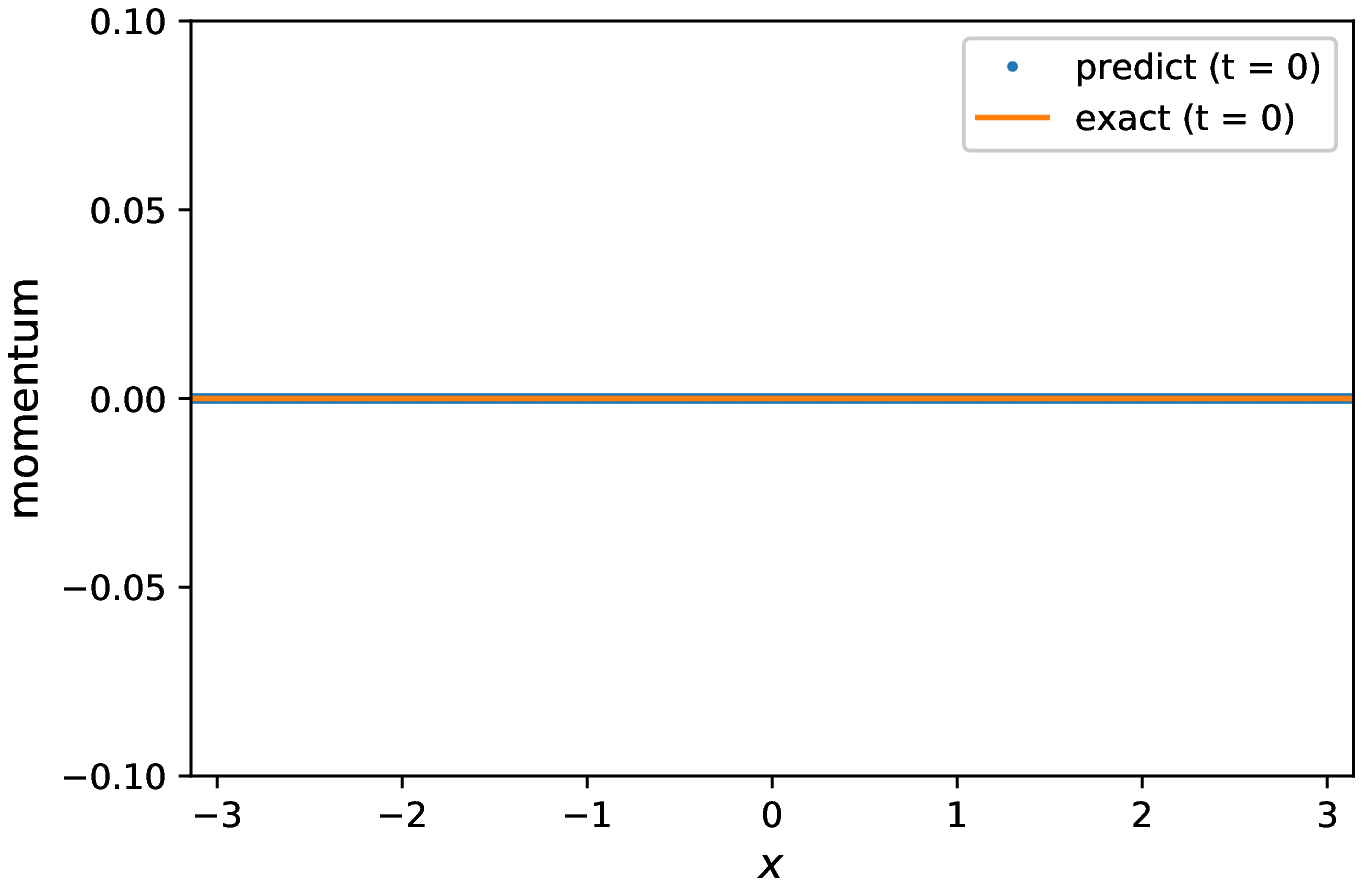}}{(ii) momentum at $t=0$}
\hspace{1cm}%
\stackunder[5pt]{\includegraphics[width=0.27\textwidth]{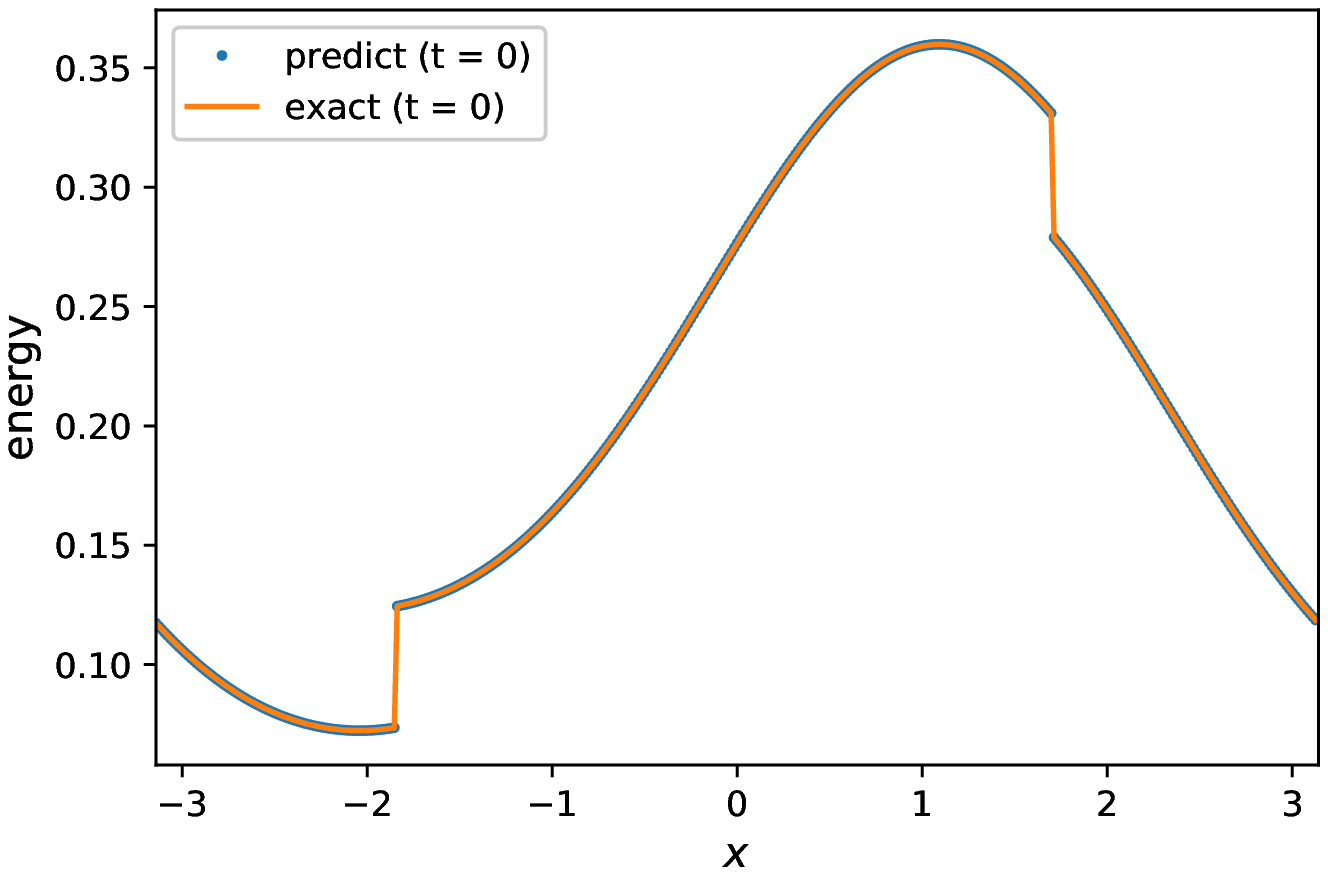}}{(iii) energy at $t=0$}
\bigskip
\stackunder[5pt]{\includegraphics[width=0.27\textwidth]{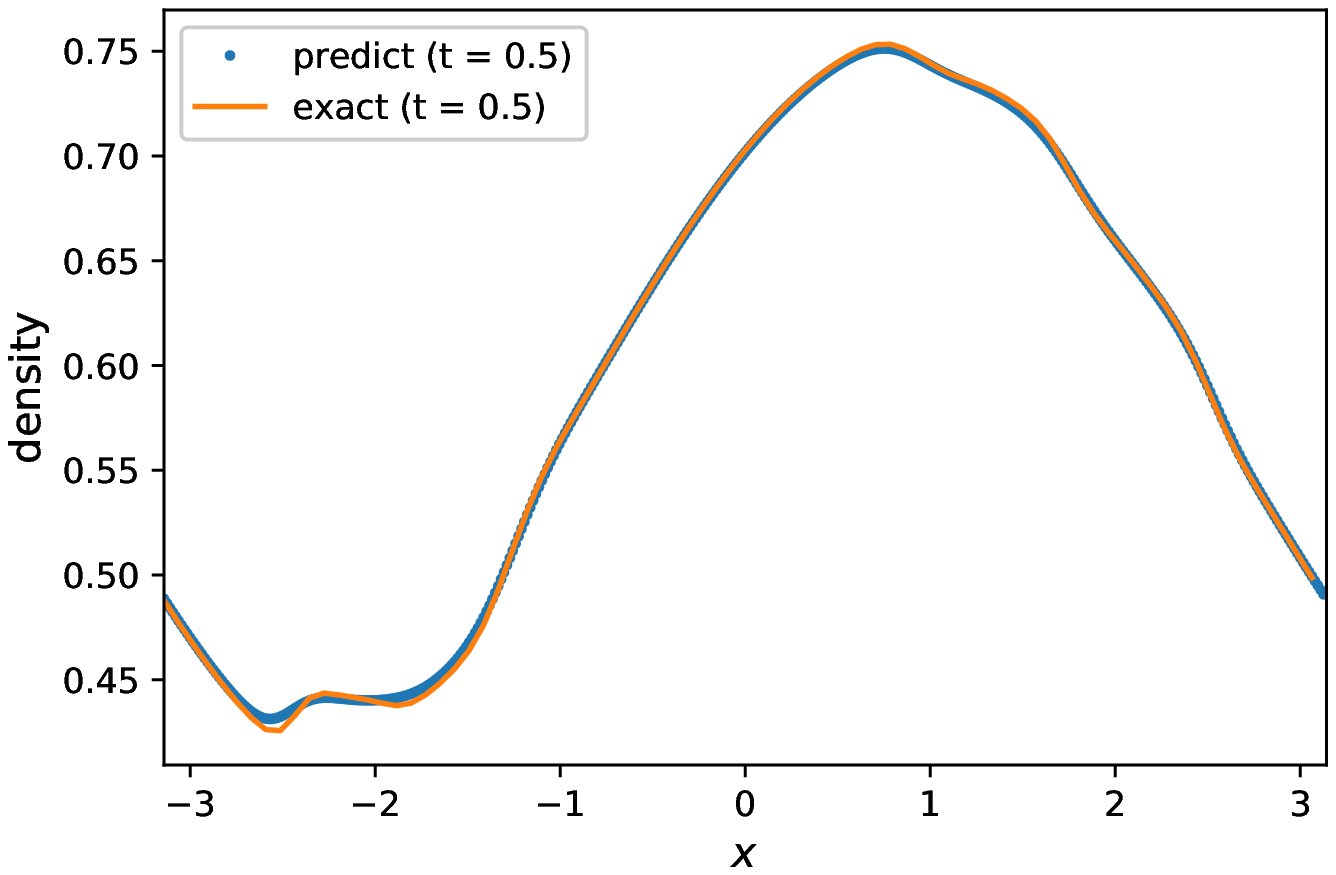}}{(i') density at $t=0.5$}
\hspace{1cm}%
\stackunder[5pt]{\includegraphics[width=0.27\textwidth]{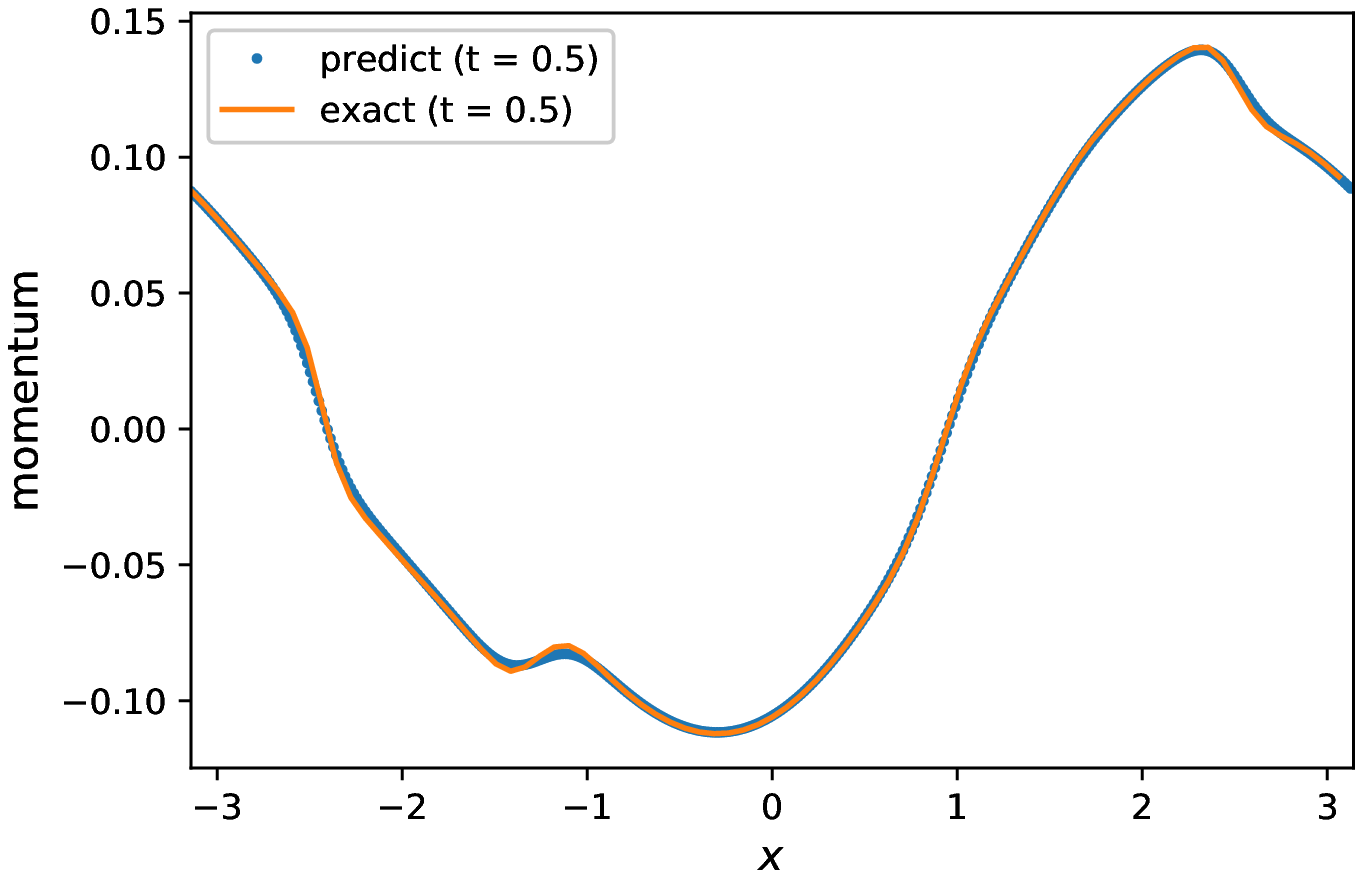}}{(ii') 
momentum at $t=0.5$}
\hspace{1cm}%
\stackunder[5pt]{\includegraphics[width=0.27\textwidth]{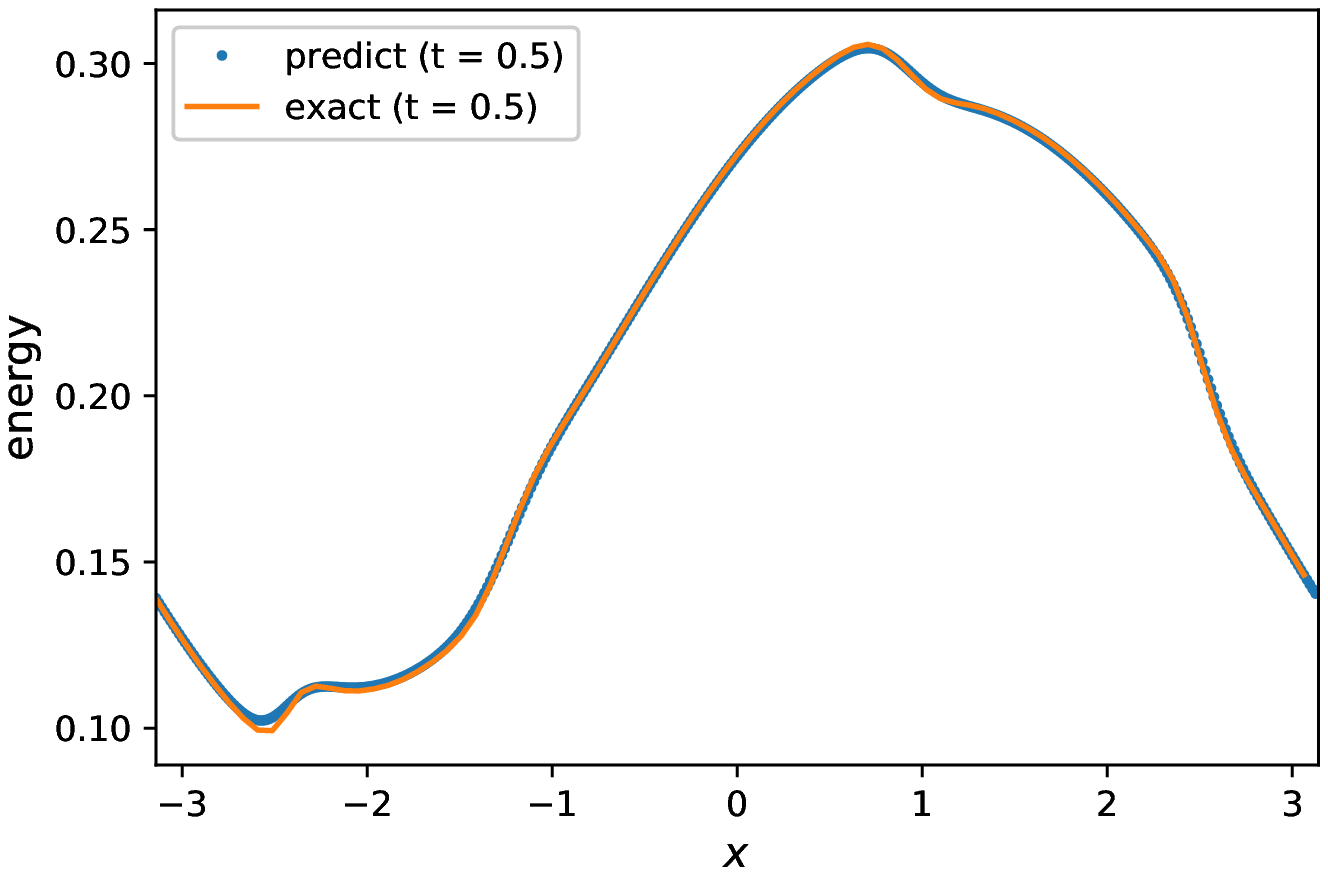}}{(iii') 
energy at $t=0.5$}
\caption{Solution profiles of density, momentum, and energy (from left to right) at $t=0$ and $t=0.5$ (from top to bottom) with $\varepsilon=10^{-2}$ and with discontinuous initial data.}
    \label{fig:shock-Kn1e-2}
\end{figure}

\begin{figure}
\footnotesize
\stackunder[5pt]{\includegraphics[width=0.27\textwidth]{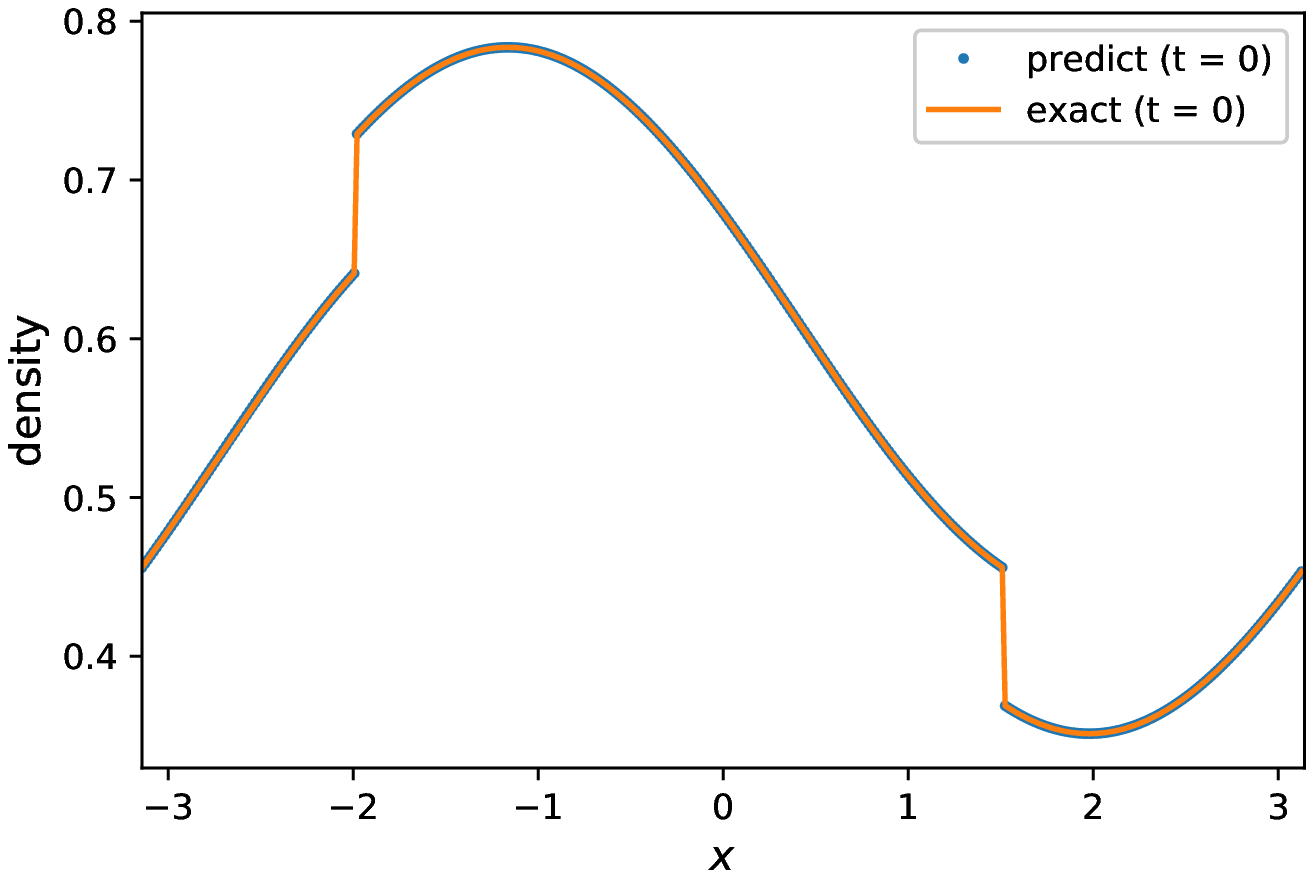}}{(i) density at $t=0$}
\hspace{1cm}%
\stackunder[5pt]{\includegraphics[width=0.27\textwidth]{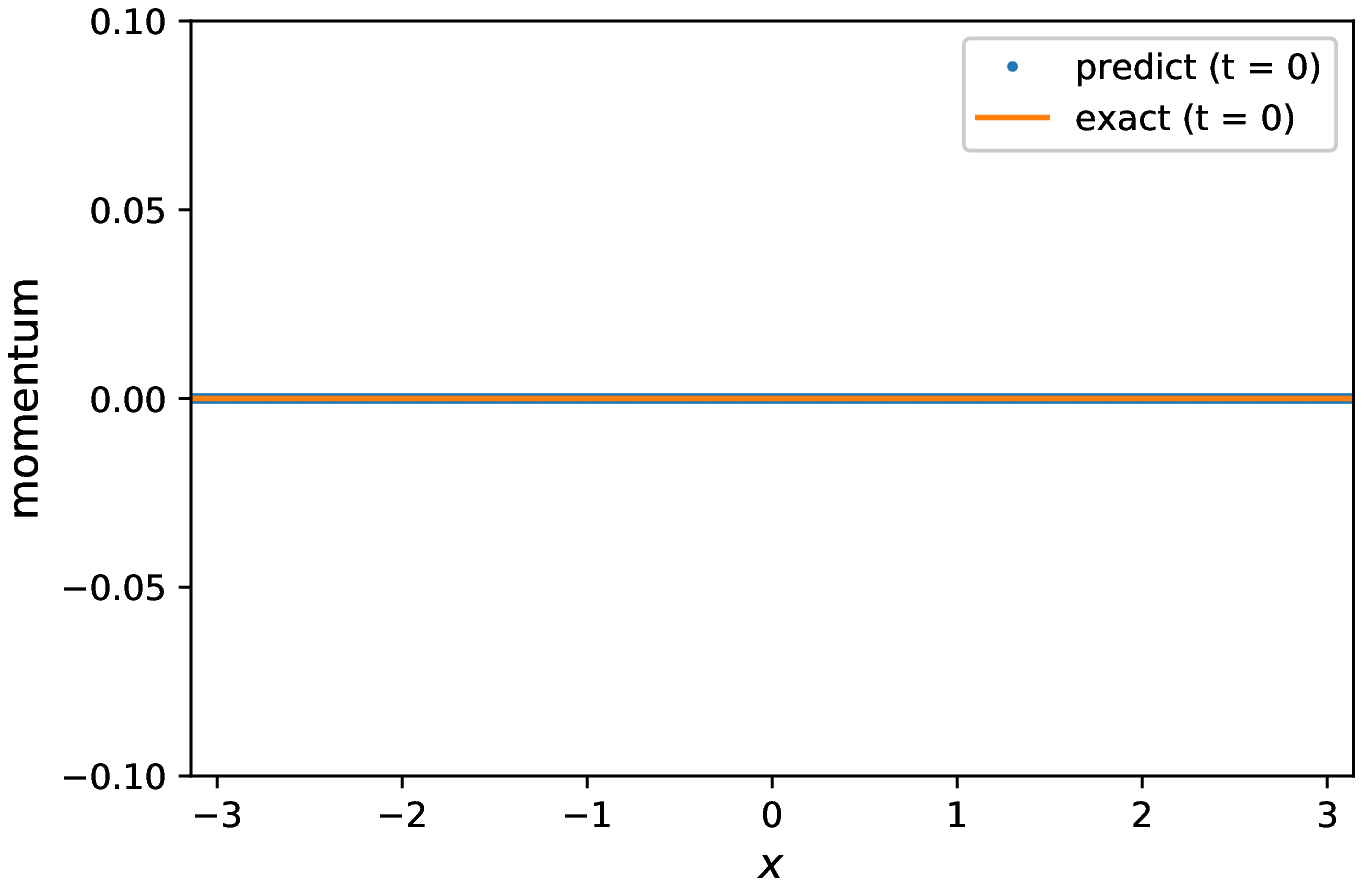}}{(ii) momentum at $t=0$}
\hspace{1cm}%
\stackunder[5pt]{\includegraphics[width=0.27\textwidth]{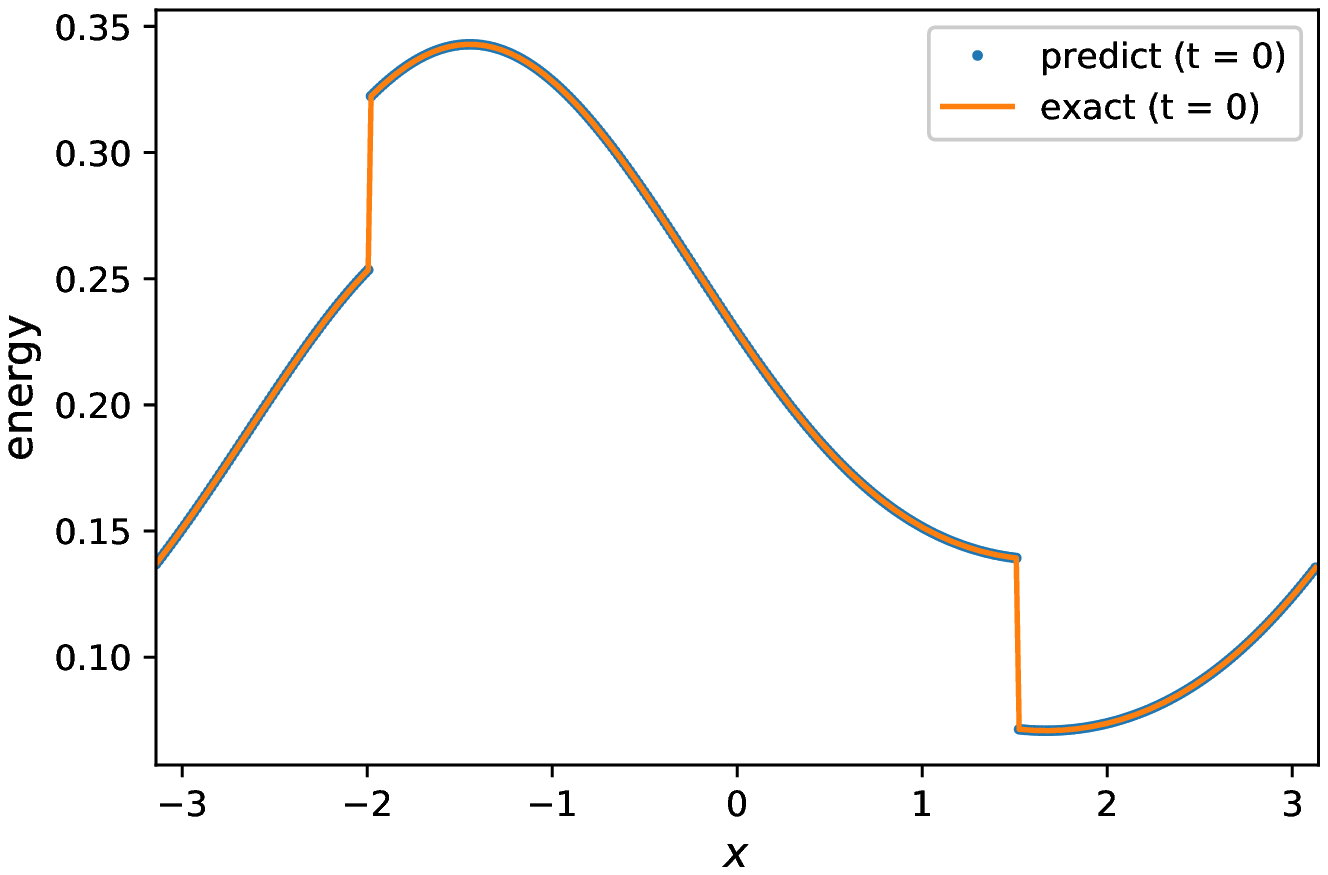}}{(iii) energy at $t=0$}
\bigskip
\stackunder[5pt]{\includegraphics[width=0.27\textwidth]{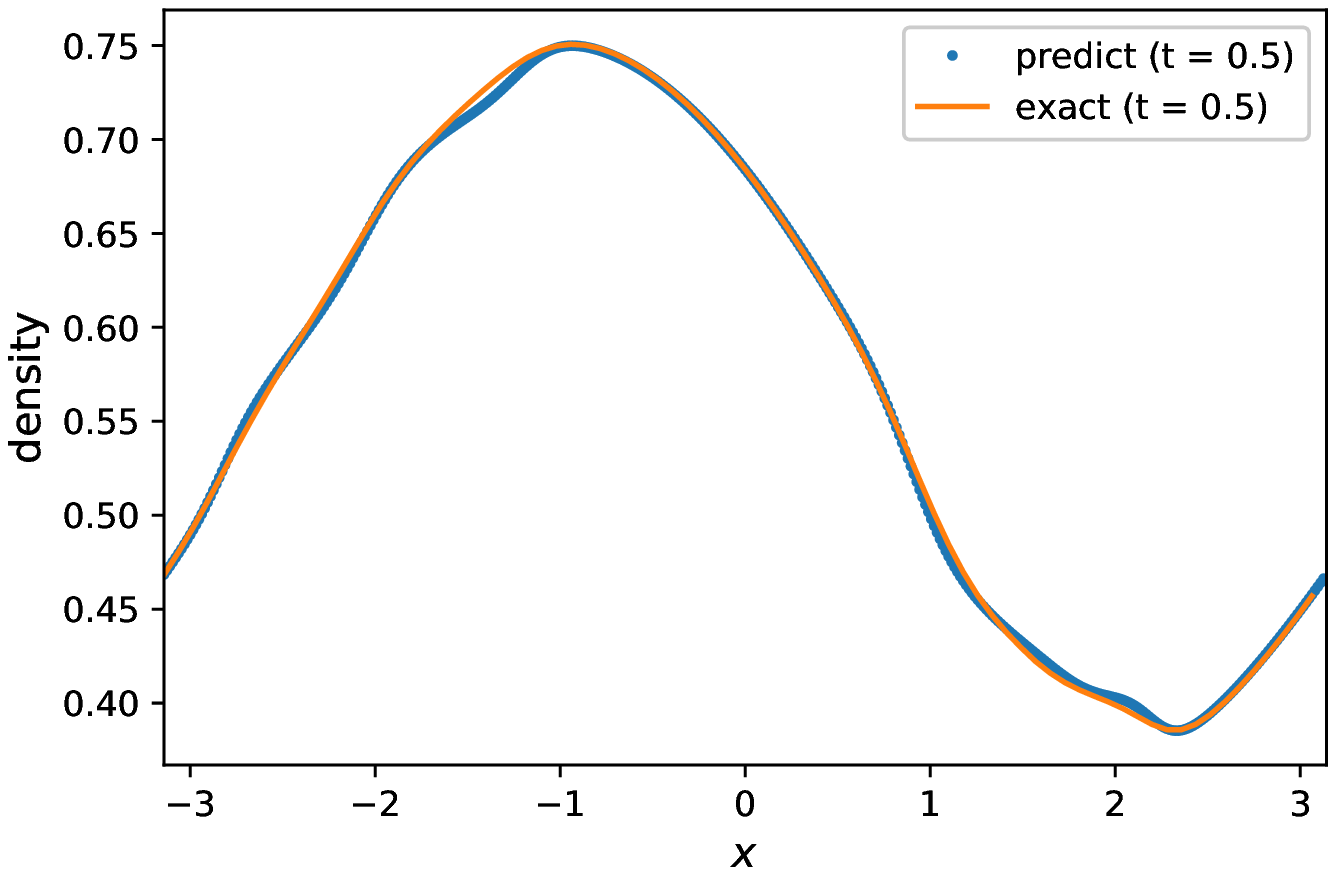}}{(i') density at $t=0.5$}
\hspace{1cm}%
\stackunder[5pt]{\includegraphics[width=0.27\textwidth]{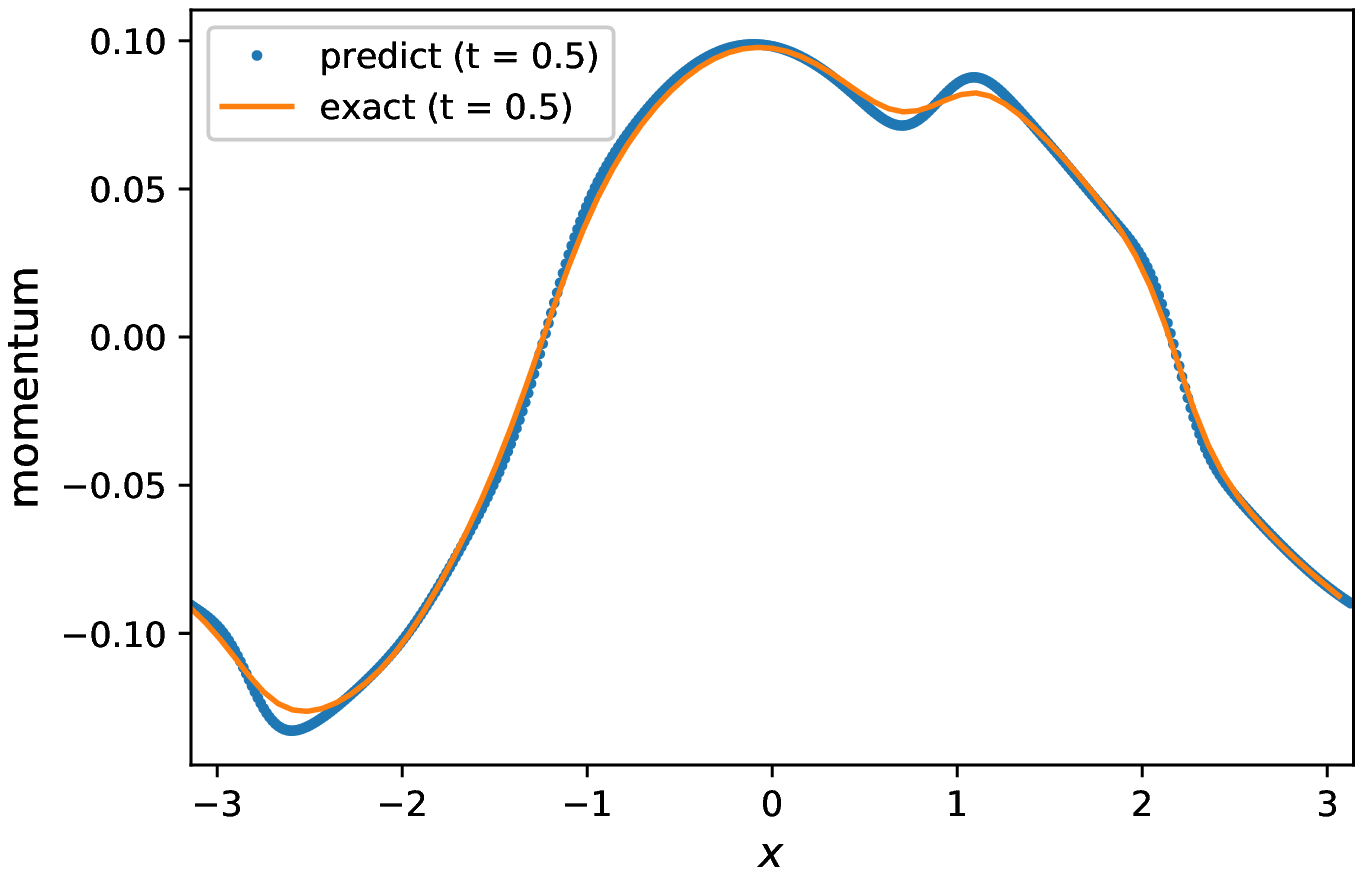}}{(ii') 
momentum at $t=0.5$}
\hspace{1cm}%
\stackunder[5pt]{\includegraphics[width=0.27\textwidth]{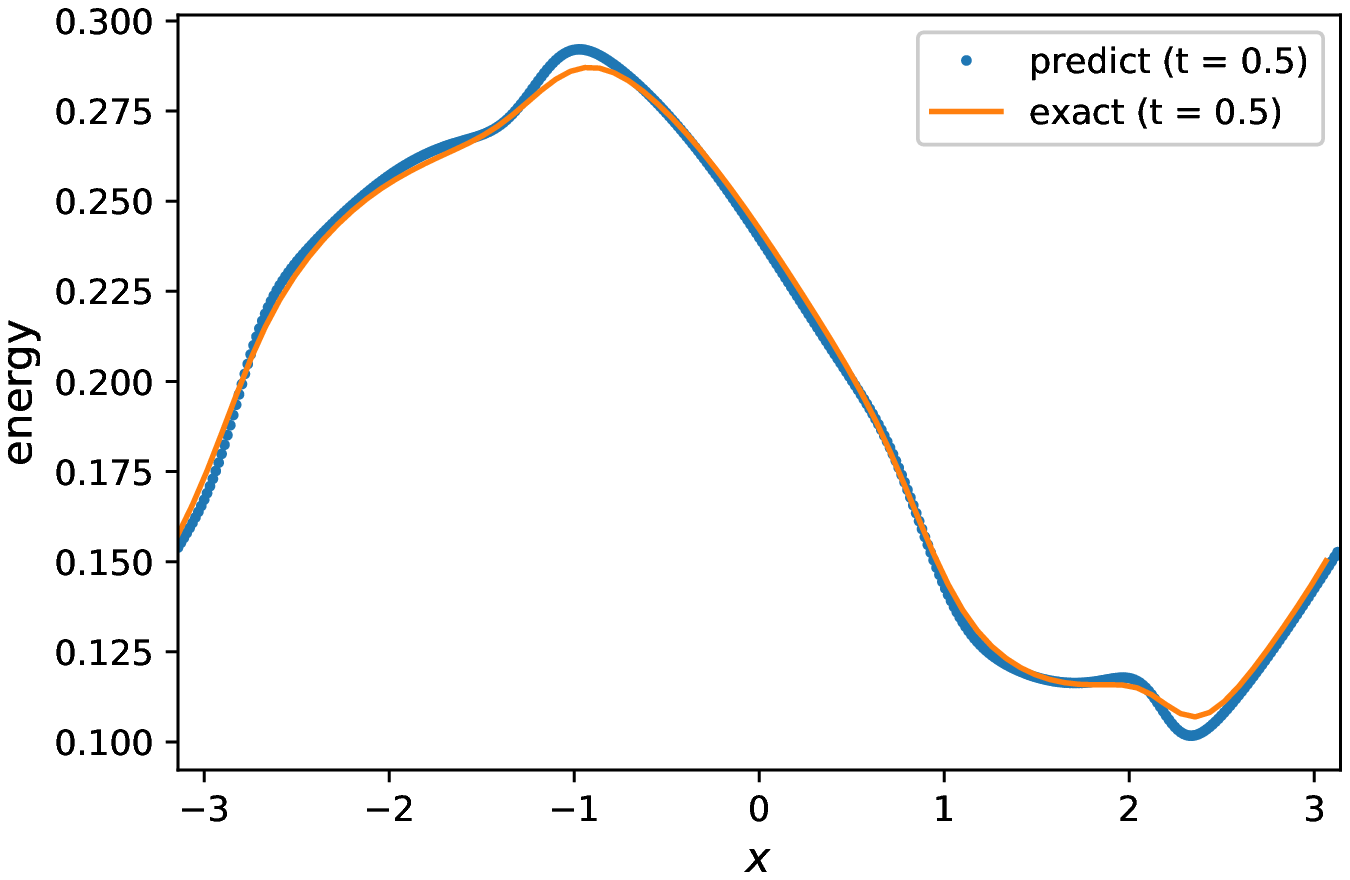}}{(iii') 
energy at $t=0.5$}
    \caption{Solution profiles of density, momentum, and energy (from left to right) at $t=0$ and $t=0.5$ (from top to bottom) with $\varepsilon=10^{-1}$ and with discontinuous initial data.}
    \label{fig:shock-Kn1e-1}
\end{figure}

\begin{figure}
\footnotesize
\stackunder[5pt]{\includegraphics[width=0.27\textwidth]{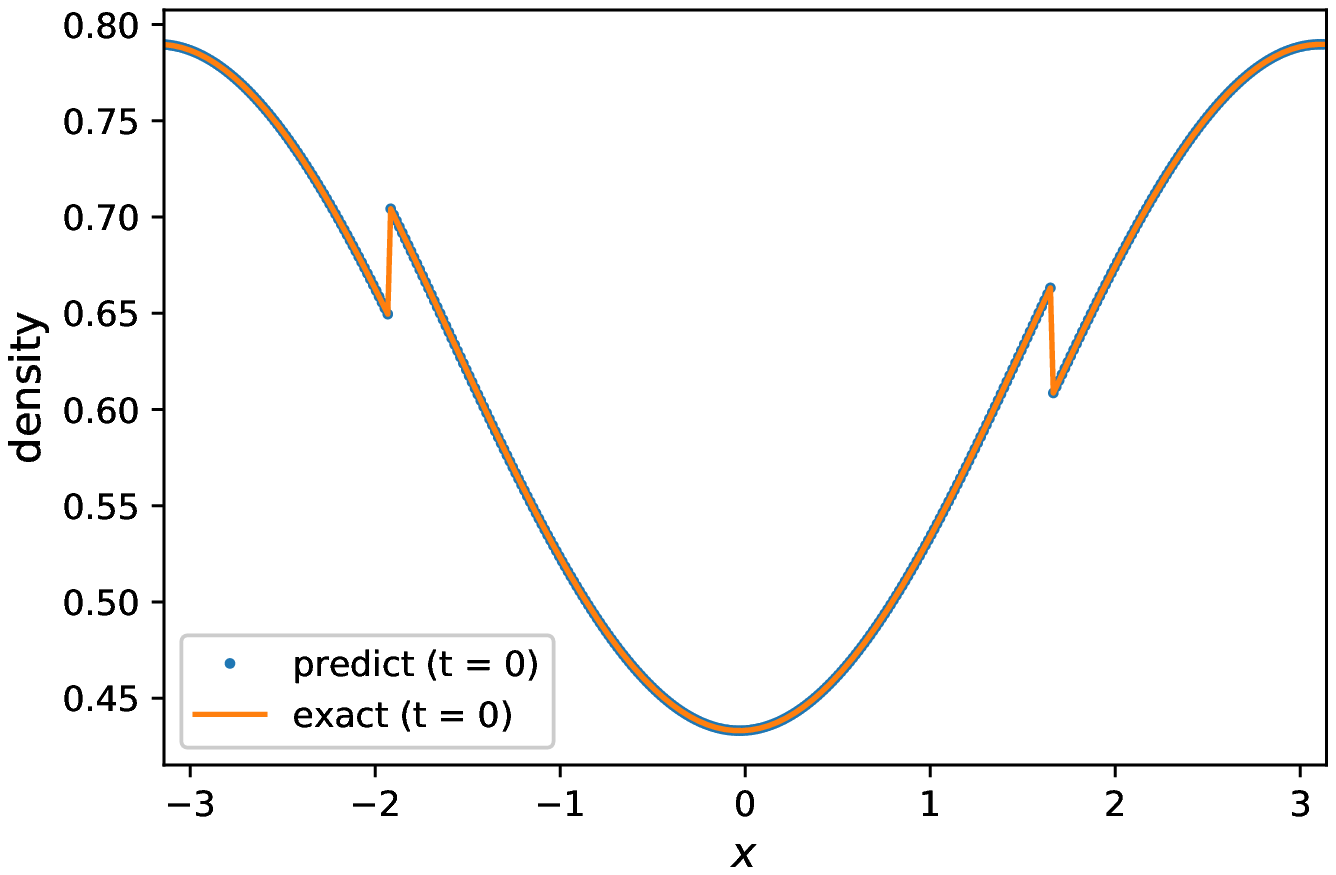}}{(i) density at $t=0$}
\hspace{1cm}%
\stackunder[5pt]{\includegraphics[width=0.27\textwidth]{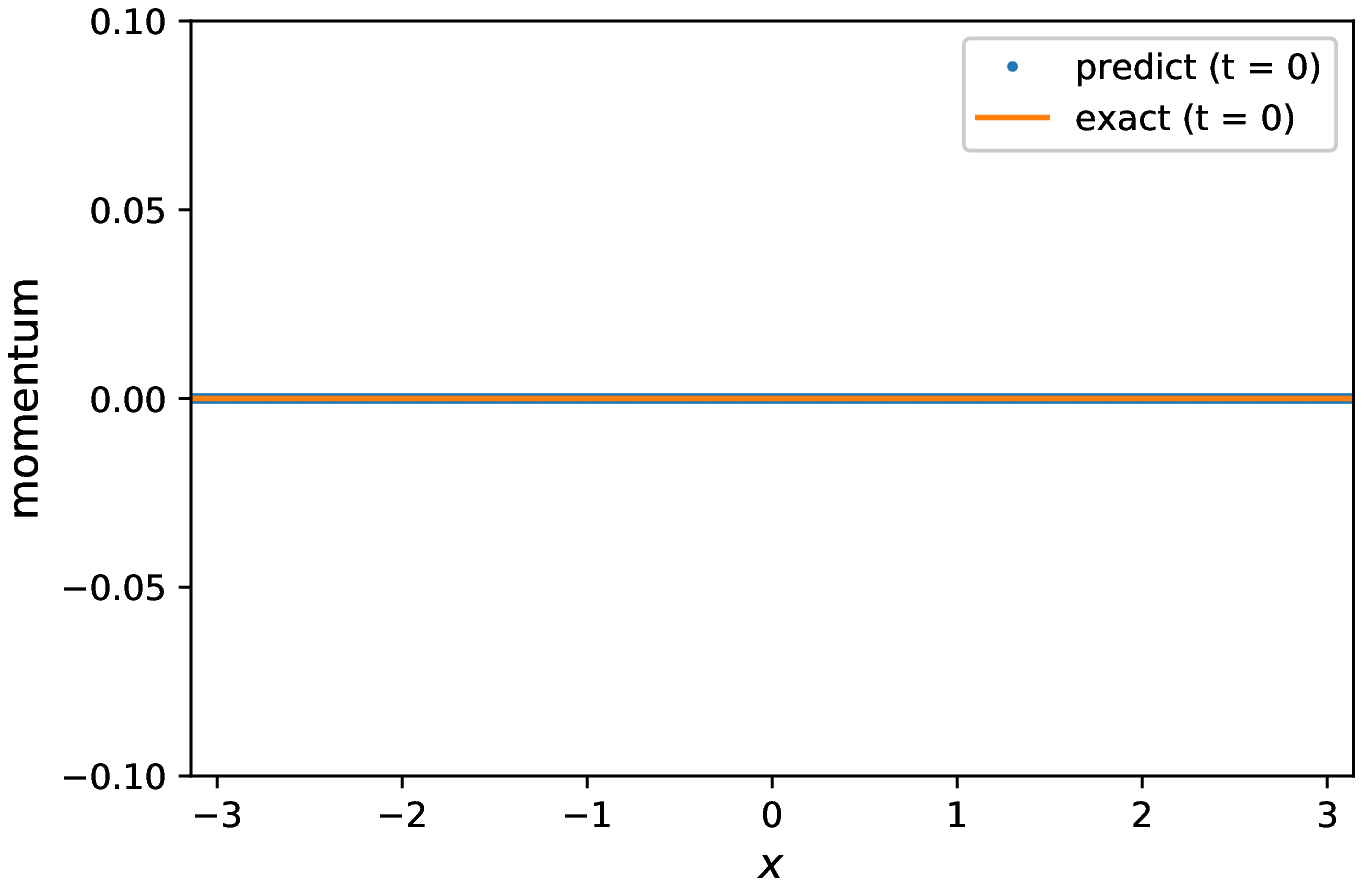}}{(ii) momentum at $t=0$}
\hspace{1cm}%
\stackunder[5pt]{\includegraphics[width=0.27\textwidth]{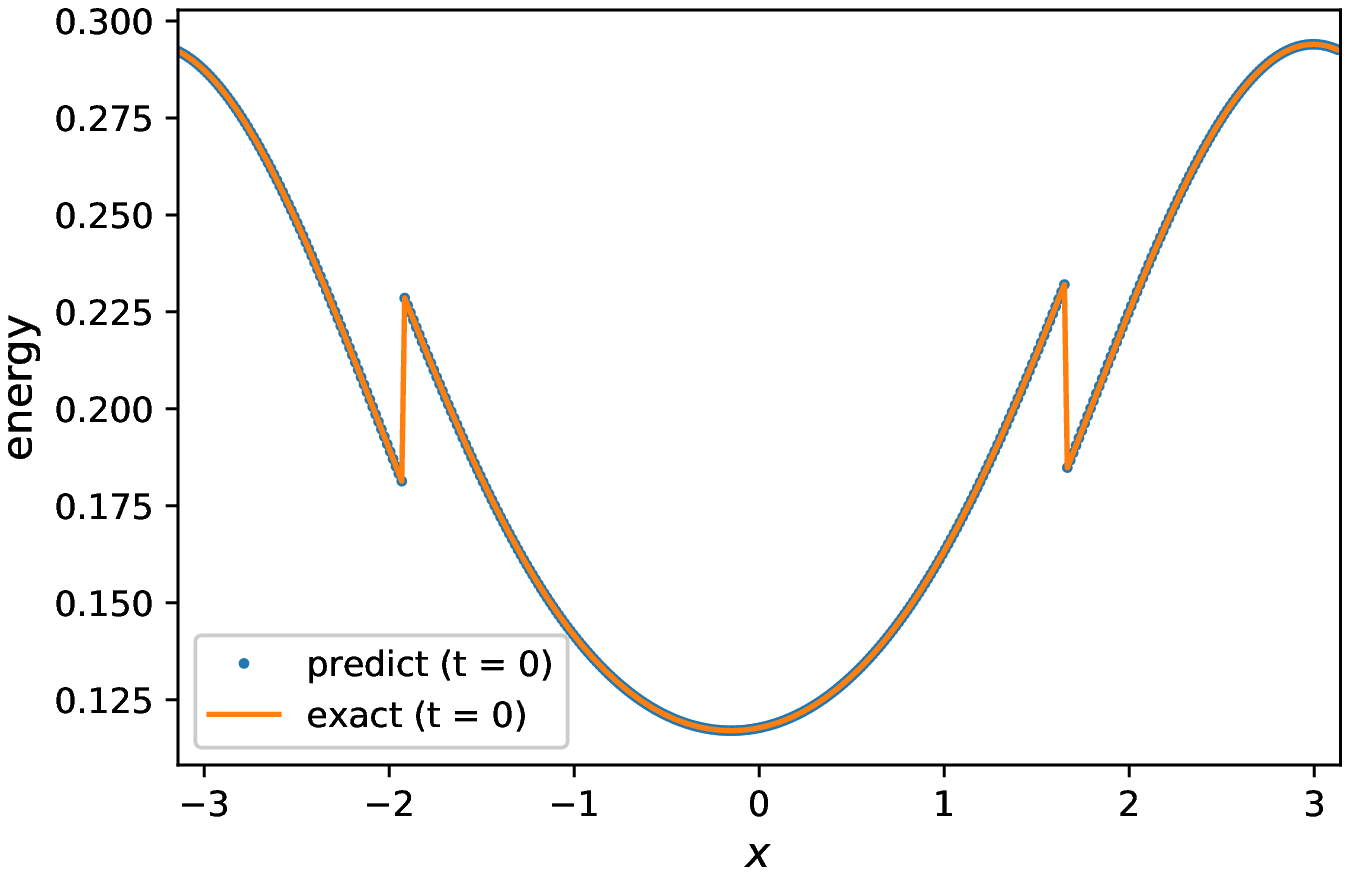}}{(iii) energy at $t=0$}
\bigskip
\stackunder[5pt]{\includegraphics[width=0.27\textwidth]{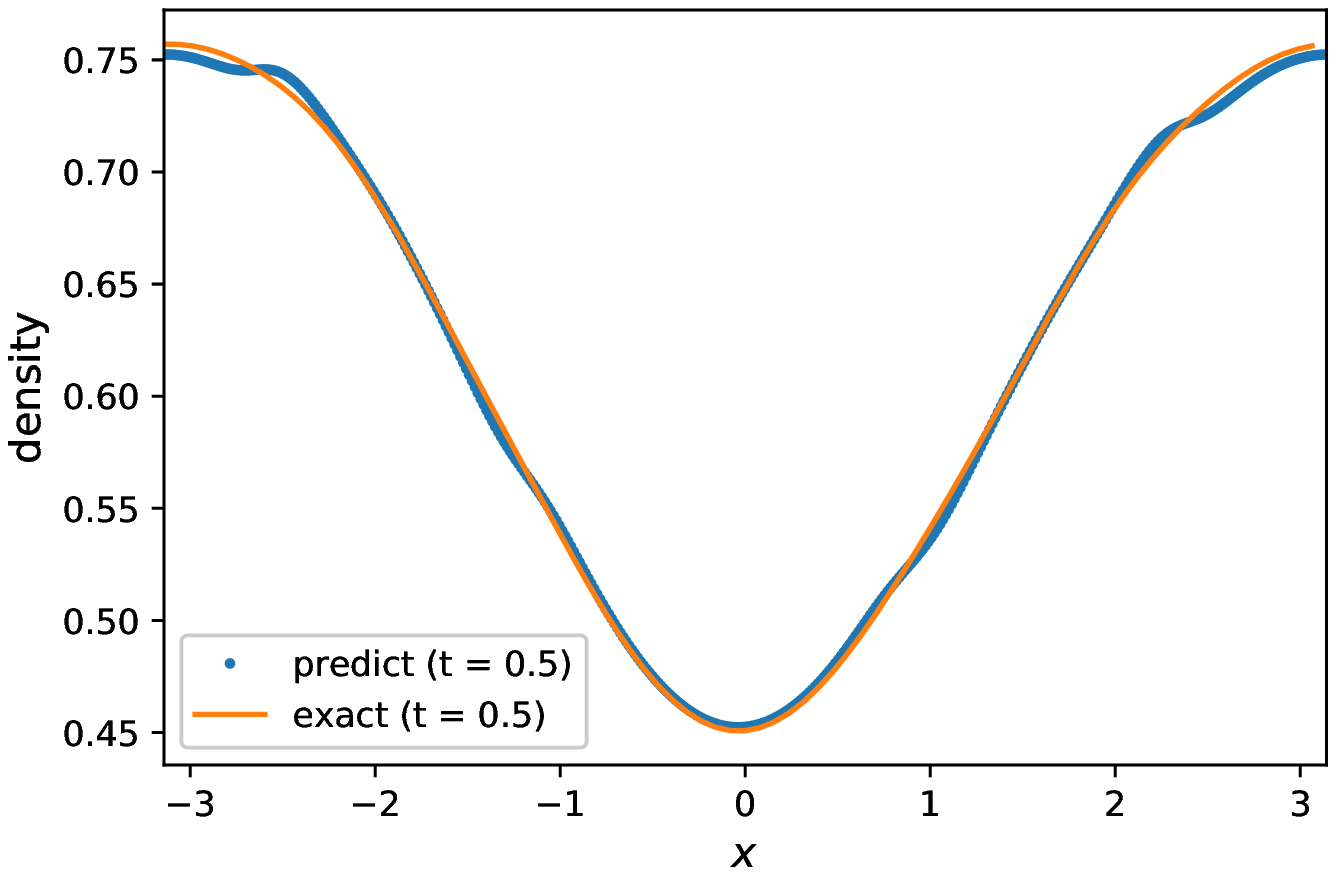}}{(i') density at $t=0.5$}
\hspace{1cm}%
\stackunder[5pt]{\includegraphics[width=0.27\textwidth]{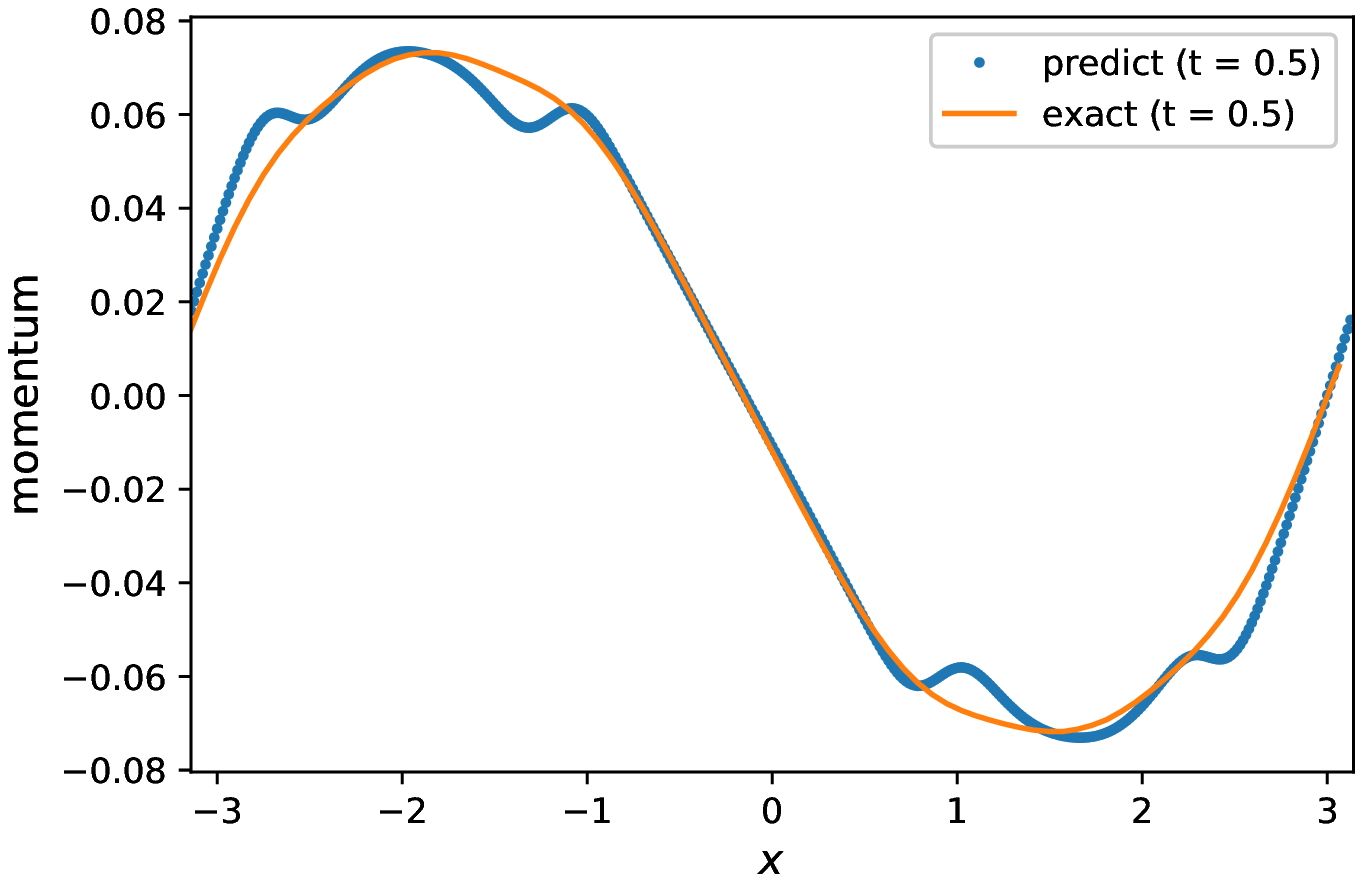}}{(ii') 
momentum at $t=0.5$}
\hspace{1cm}%
\stackunder[5pt]{\includegraphics[width=0.27\textwidth]{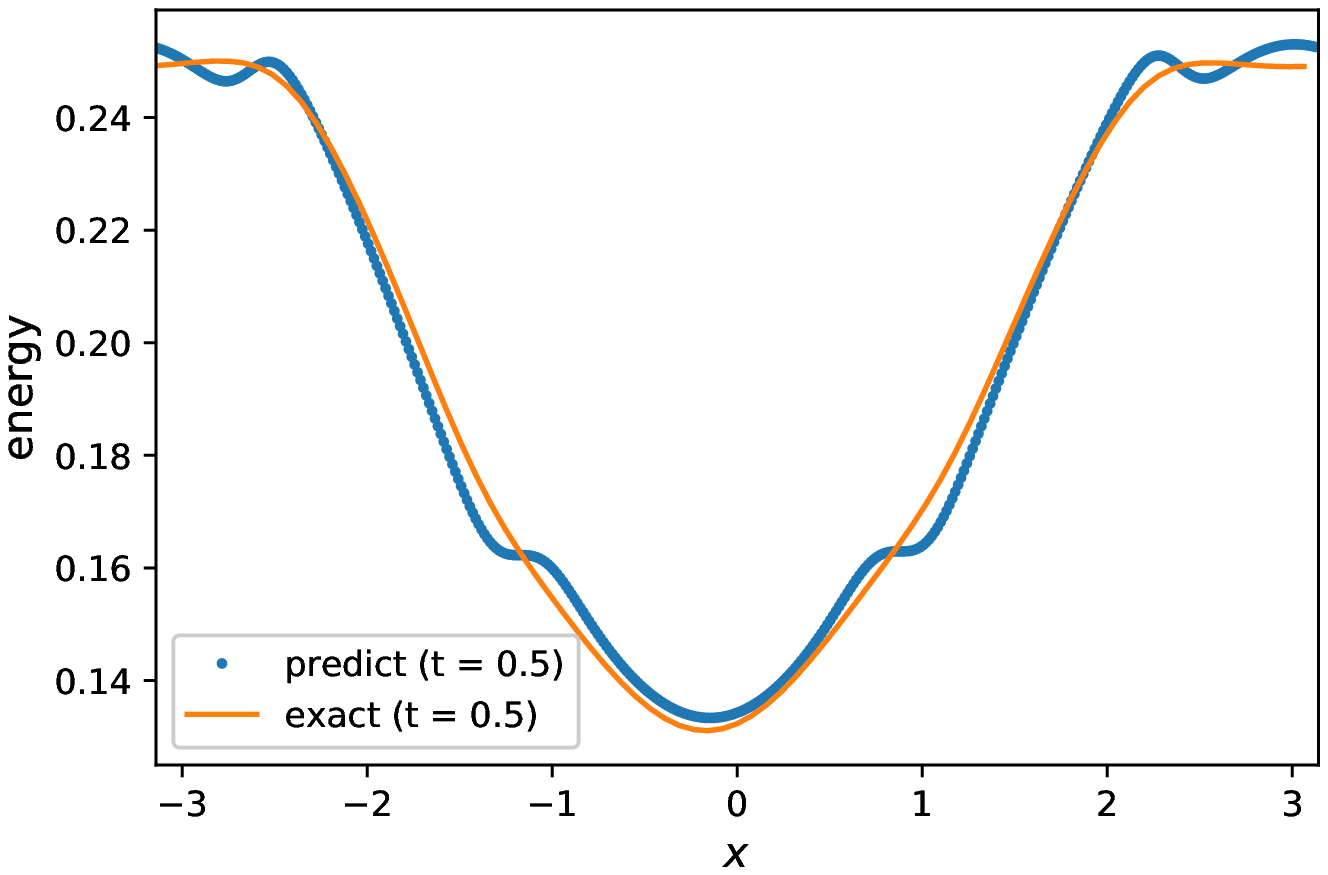}}{(iii') 
energy at $t=0.5$}
    \caption{Solution profiles of density, momentum, and energy (from left to right) at $t=0$ and $t=0.5$ (from top to bottom) with $\varepsilon=1$ and with discontinuous initial data.}
    \label{fig:shock-Kn1e0}
\end{figure}

\begin{figure}
\footnotesize
\stackunder[5pt]{\includegraphics[width=0.27\textwidth]{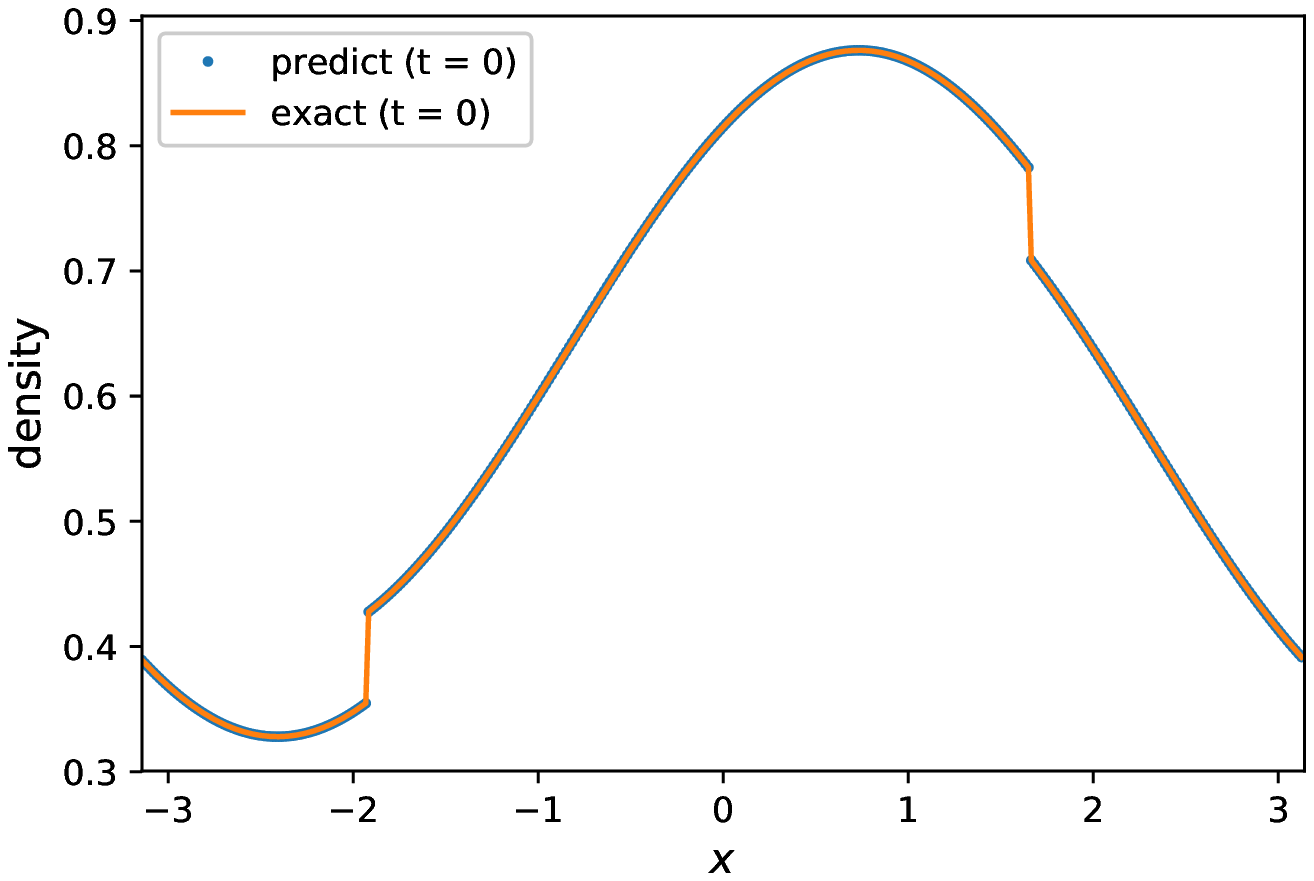}}{(i) density at $t=0$}
\hspace{1cm}%
\stackunder[5pt]{\includegraphics[width=0.27\textwidth]{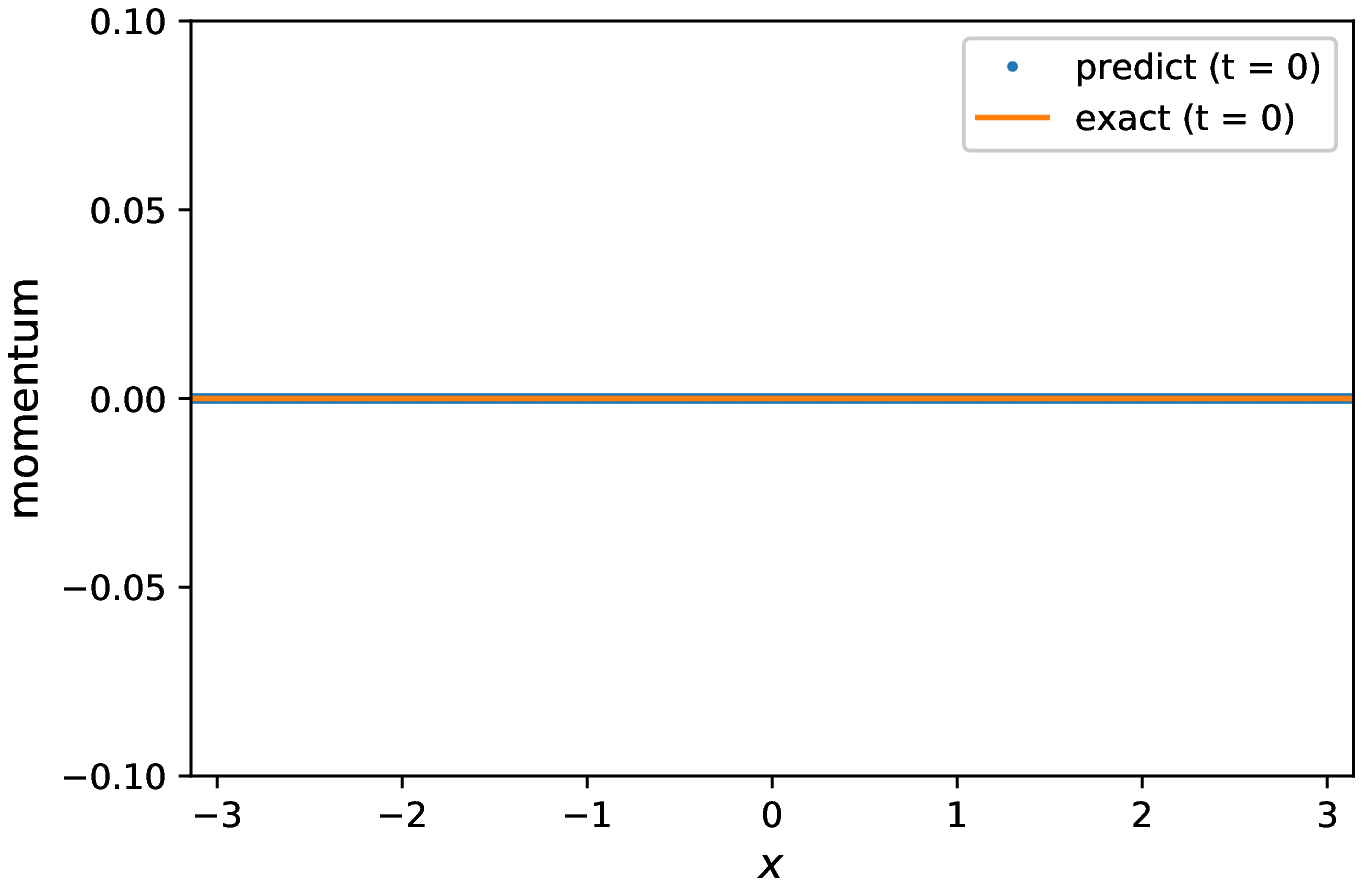}}{(ii) momentum at $t=0$}
\hspace{1cm}%
\stackunder[5pt]{\includegraphics[width=0.27\textwidth]{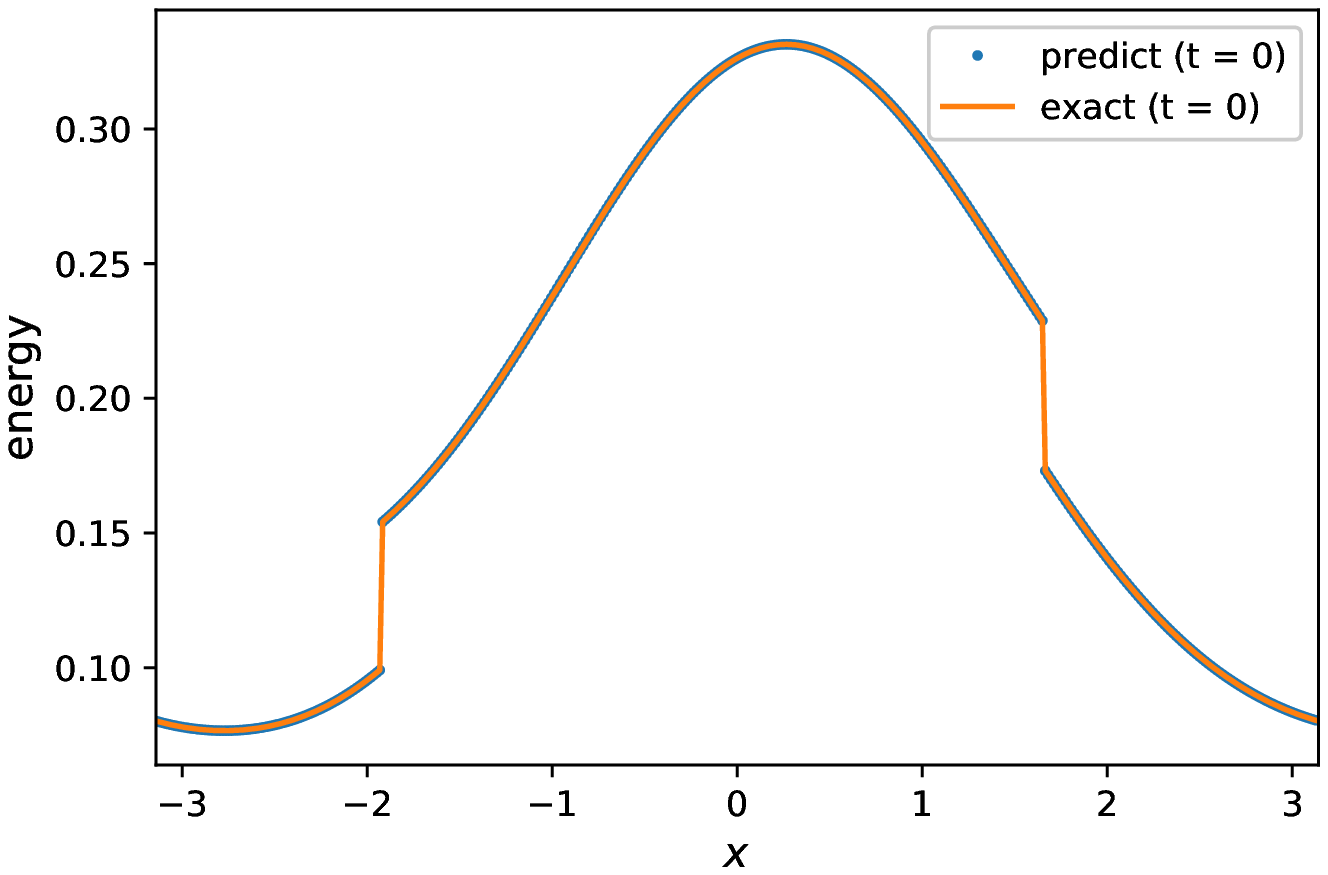}}{(iii) energy at $t=0$}
\bigskip
\stackunder[5pt]{\includegraphics[width=0.27\textwidth]{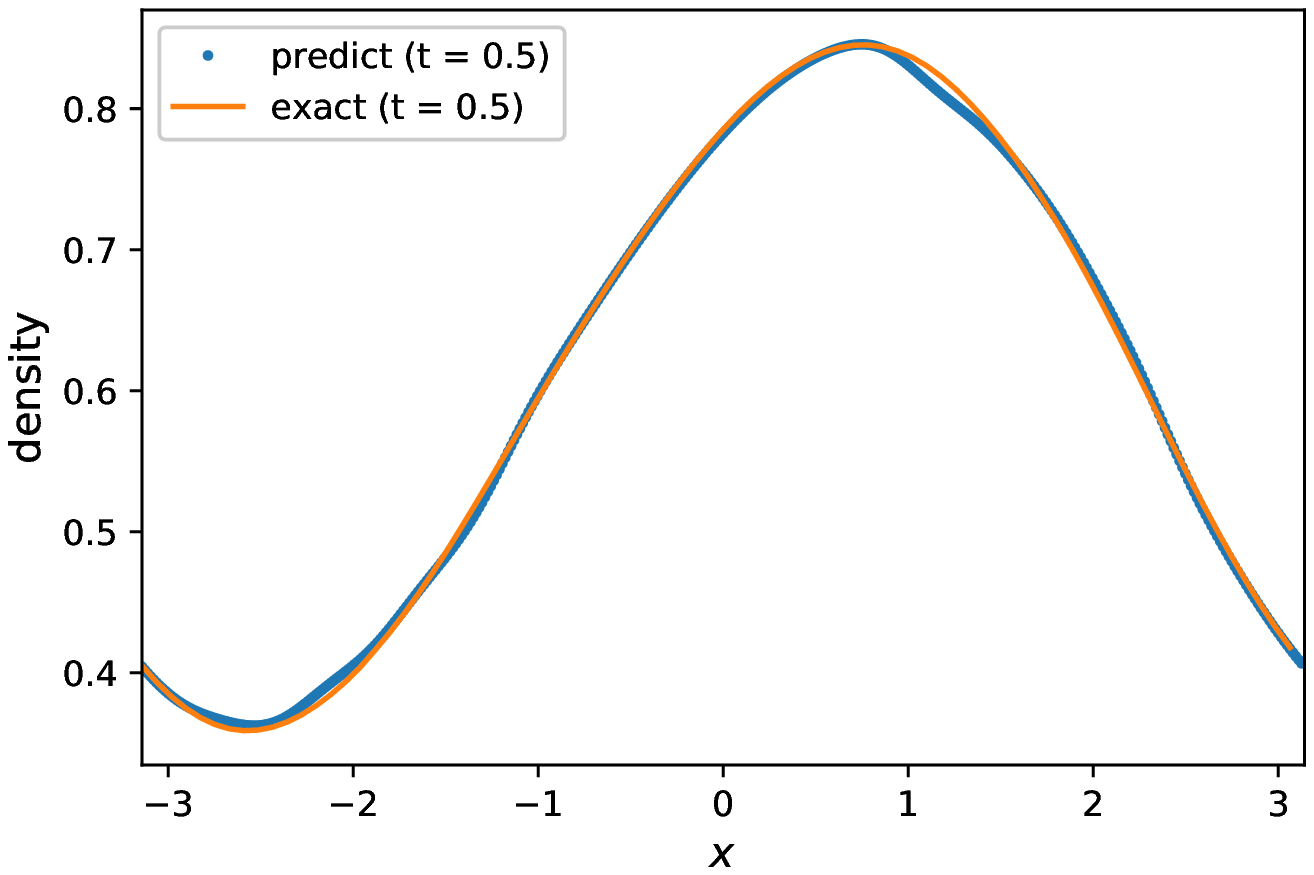}}{(i') density at $t=0.5$}
\hspace{1cm}%
\stackunder[5pt]{\includegraphics[width=0.27\textwidth]{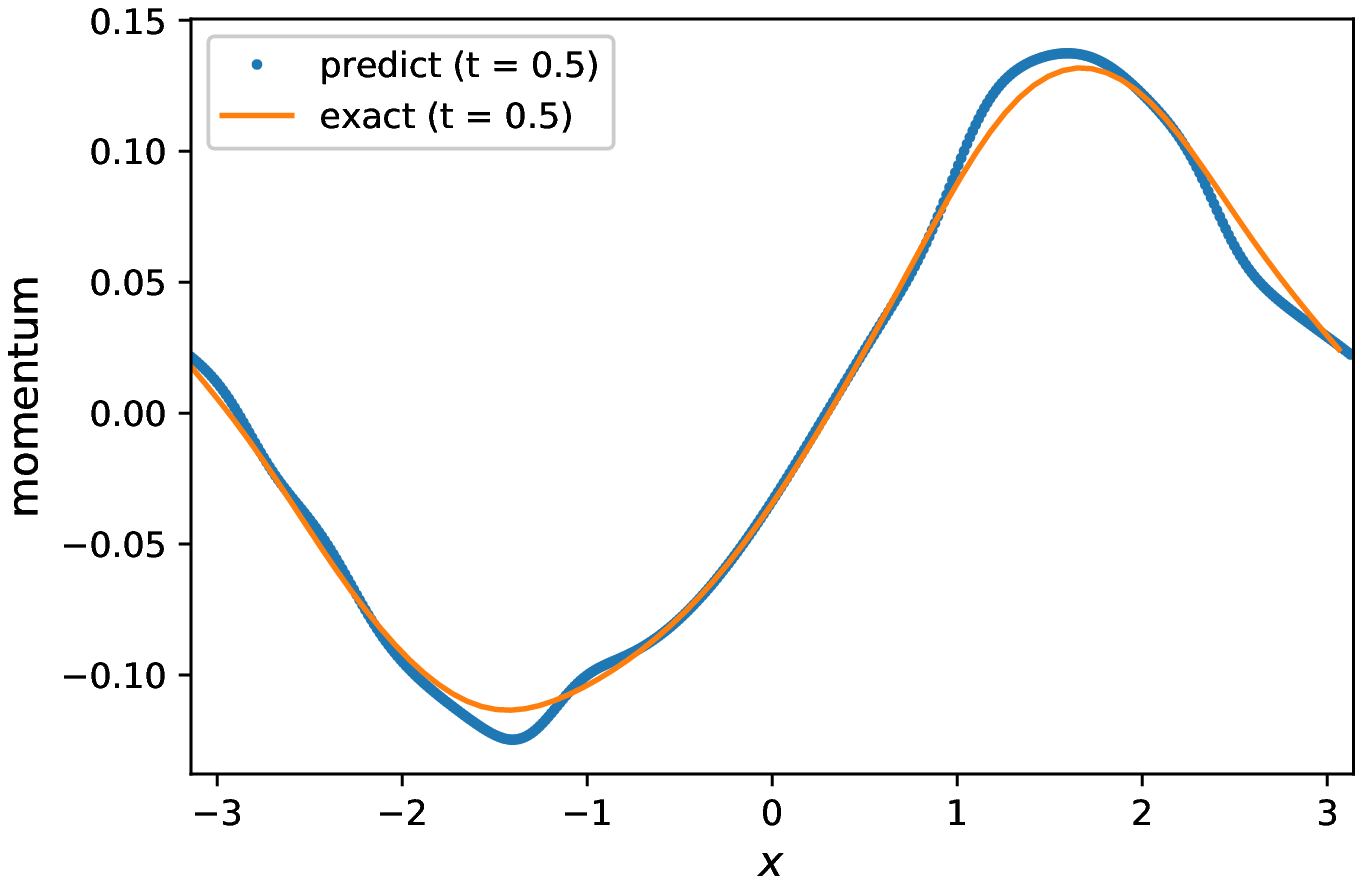}}{(ii') 
momentum at $t=0.5$}
\hspace{1cm}%
\stackunder[5pt]{\includegraphics[width=0.27\textwidth]{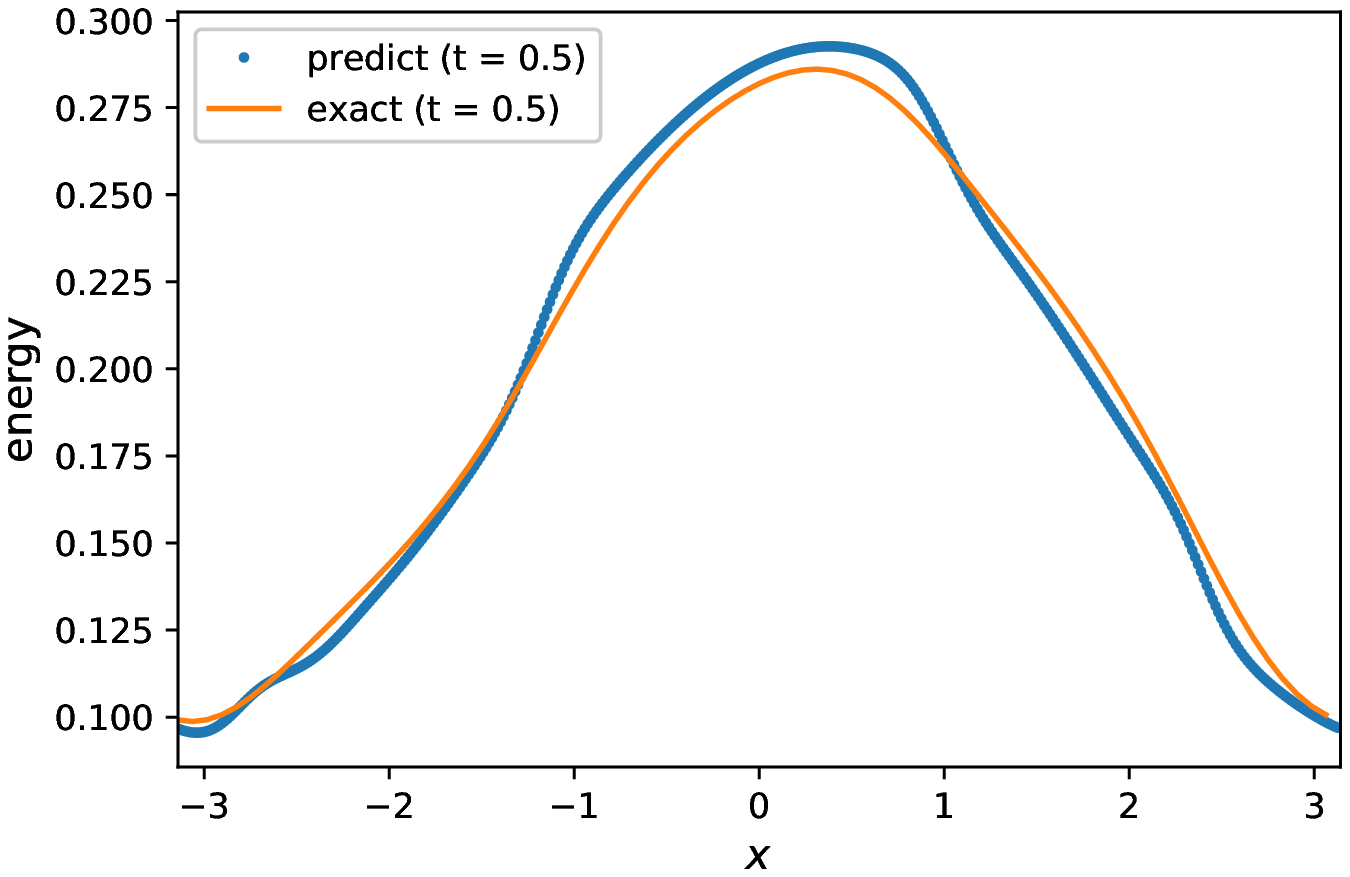}}{(iii') 
energy at $t=0.5$}
    \caption{Solution profiles of density, momentum, and energy (from left to right) at $t=0$ and $t=0.5$ (from top to bottom) with $\varepsilon=10$ and with discontinuous initial data.}
    \label{fig:shock-Kn1e1}
\end{figure}

Below is the solution profiles with discontinuous initial data. 
Figures \ref{fig:shock-Kn1e-3}-\ref{fig:shock-Kn1e1} are for different Knudsen numbers $\varepsilon=10^{-3}, 10^{-2}, 10^{-1}, 1$ and $10$.
For small Knudsen numbers ($\varepsilon=10^{-3}, 10^{-2}, 1$), our learned model agrees quite well with the BGK model, see Figures \ref{fig:shock-Kn1e-3}-\ref{fig:shock-Kn1e-1}. 
However, the deviation becomes large for $\varepsilon=1$ and 10, see Figures \ref{fig:shock-Kn1e0} and \ref{fig:shock-Kn1e1}. 

\begin{figure}
	\footnotesize
	\stackunder[5pt]{\includegraphics[width=0.27\textwidth]{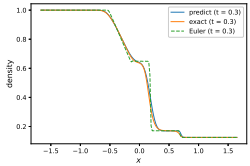}}{(i) density at $t=0.3$}
	\hspace{1cm}%
	\stackunder[5pt]{\includegraphics[width=0.27\textwidth]{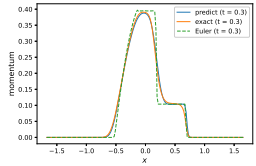}}{(ii) 
	momentum at $t=0.3$}
	\hspace{1cm}%
	\stackunder[5pt]{\includegraphics[width=0.27\textwidth]{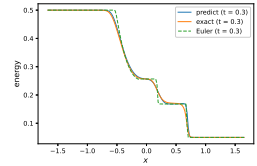}}{(iii) 
	energy at $t=0.3$}
	\caption{Sod shock tube problem. Solution profiles of density, momentum, and energy (from left to right) at $t=0.3$ with $\varepsilon=10^{-2}$.}
    \label{fig:sod-Kn1e-2}
\end{figure}

\begin{figure}
	\footnotesize
	\stackunder[5pt]{\includegraphics[width=0.27\textwidth]{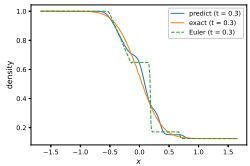}}{(i) density at $t=0.3$}
	\hspace{1cm}%
	\stackunder[5pt]{\includegraphics[width=0.27\textwidth]{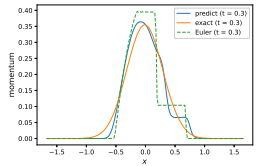}}{(ii) 
	momentum at $t=0.3$}
	\hspace{1cm}%
	\stackunder[5pt]{\includegraphics[width=0.27\textwidth]{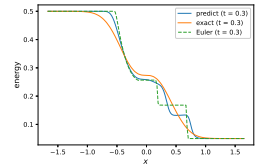}}{(iii) 
	energy at $t=0.3$}
	\caption{Sod shock tube problem. Solution profiles of density, momentum, and energy (from left to right) at $t=0.3$ with $\varepsilon=1$.}
    \label{fig:sod-Kn1e0}
\end{figure}

{

We further examine our model on Sod's shock tube problem \cite{sod1978survey} with initial data
\begin{equation}\label{eq:sod-init}
(\rho, v, p)_{\textrm{}}(x,0) = 
\left\{
\begin{aligned}
& (1, 0, 1), \quad &x<0, \\
& (0.125, 0, 0.1), \quad &x>0. \\
\end{aligned}
\right.
\end{equation}
With these initial data, we numerically solve our model, the Euler equations and the kinetic equation up to $t=0.3$. 
In the computations of our learned equations and the Euler equations, we use the Lax-Friedrich scheme with $N_x=1600$.
The solution profiles of density, momentum, and energy obtained from the three models are shown in Figure \ref{fig:sod-Kn1e-2} and Figure \ref{fig:sod-Kn1e0} with Knudsen number $\varepsilon=10^{-2}$ and $\varepsilon=1$.

From Figure \ref{fig:sod-Kn1e-2} with $\varepsilon=10^{-2}$, it is observed that the solution profiles of our model agree quite well with those of the BGK model. In contrast, the Euler equations generate sharp discontinuity, which deviates from the BGK model. 
On the other hand, Figure \ref{fig:sod-Kn1e0} is for larger Knudsen number $\varepsilon=1$. In this case, it is far away from equilibrium and difficult to build a good macroscopic model. We observe that the solutions of the Euler equations are quite different from those of the kinetic model, while our learned model also has a slight discrepancy but can well capture the tendency of the solution. The discrepancy may be due to the fact that only one non-equilibrium variable $w$ is introduced in our model. The poor behavior was also observed for the moment closure systems with few number of moments and improved with larger number of moments \cite{cai2014dimension,cai2010numerical}. Thus, we expect that the discrepancy could be reduced by adding more non-equilibrium variables under the CDF framework and this is left for future work.
}

\section{Concluding remarks}\label{sec:conclusion}

In this work, we advocate combining machine learning with non-equilibrium thermodynamics. Precisely, we develop a method for learning interpretable, thermodynamically stable and Galilean invariant PDEs based on CDF. As governing equations for one-dimensional non-equilibrium flows, the learned PDEs are parameterized by fully-connected neural networks and satisfy the conservation-dissipation principle automatically. In particular, they are hyperbolic balance laws.

The training data are generated from the BGK model with smooth initial data.  
Numerical results indicate that our CDF-based machine learning model achieves good accuracy in a wide range of Knudsen numbers.
It is remarkable that the learned dynamics can give satisfactory results with randomly sampled discontinuous initial data although it is trained only with smooth initial data. {Particularly, for the classical Sod's shock tube problem, our model behaves much better than the Euler equations.}

This work could be improved in the following aspects: (1) Our model has only one non-equilibrium variable which may not be the best. Better numerical results are expected by introducing more non-equilibrium variables. (2) Only one-dimensional problems are considered here. It is more challenging to generalize our method to problems in multi-dimension.

\section*{Acknowledgment}

JH would like to thank Qi Tang in Los Alamos National Laboratory for helpful discussions in training of the neural networks.

\bibliographystyle{abbrv}
\bibliography{ref}

\end{document}